\documentclass[iop,revtex4,numberedappendix,twocolappendix]{emulateapj}
\usepackage{times,epsfig} 
\usepackage{natbib}
\usepackage{amssymb,amsmath}
\usepackage{graphicx}
\usepackage{lscape}

\newcommand{\nms}{\mathsurround=0pt}
\newcommand{\oversim}[2]{\lower 2pt\vbox{\baselineskip 0pt \lineskip 1pt \ialign{$\nms#1\hfil##\hfil$\crcr#2\crcr\sim\crcr}}}
\newcommand{\gtsim}{\mathrel{\mathpalette\oversim{>}}}
\newcommand{\ltsim}{\mathrel{\mathpalette\oversim{<}}}
\newcommand{\HI}{H{\sc I}~}
\newcommand{\HIns}{H{\sc I}}
\newcommand{\HII}{H{\sc II}~}

\newcommand{\NH}{$N_{\mathrm{H}}$~}
\newcommand{\NHI}{$N_{\mathrm{HI}}$~}
\newcommand{\NHns}{$N_{\mathrm{H}}$}
\newcommand{\NHIns}{$N_{\mathrm{HI}}$}

\newcommand{\NHIIns}{$N_{\mathrm{HII}}$}
\newcommand{\persec}{s$^{-1}\,$}
\newcommand{\percmq}{cm$^{-2}\,$}

\newcommand{\perhz}{Hz$^{-1}$}
\newcommand{\tnm}{\tablenotemark}

\shorttitle{\NH - \NHI correlation in compact radio galaxies}
\shortauthors{Ostorero et al.}

\begin{document}

\title{Correlation between X-ray and radio absorption in compact radio galaxies}

\author{Luisa Ostorero\altaffilmark{1,2}}\email{ostorero@ph.unito.it}
\author{Raffaella Morganti\altaffilmark{3,4}}
\author{Antonaldo Diaferio\altaffilmark{1,2}}
\author{Aneta Siemiginowska\altaffilmark{5}}
\author{\\{\L}ukasz Stawarz\altaffilmark{6}} 
\author{Rafal Moderski\altaffilmark{7}} 
\author{Alvaro Labiano\altaffilmark{8}}

\altaffiltext{1}{Dipartimento di Fisica, Universit\`a degli 
       Studi di Torino, Via P. Giuria 1, I-10125 Torino, Italy}
\altaffiltext{2}{Istituto Nazionale di Fisica Nucleare (INFN), Sezione di Torino, Via 
       P. Giuria 1, I-10125 Torino, Italy}
\altaffiltext{3}{Netherlands Institute for Radio Astronomy, Postbus 2, 7990 AA Dwingeloo, 
       The Netherlands}
\altaffiltext{4}{Kapteyn Astronomical Institute, University of Groningen, P.O. Box 800, 
       9700 AV Groningen, The Netherlands}
\altaffiltext{5}{Harvard-Smithsonian Center for Astrophysics, 60 Garden St., Cambridge, MA 02138, USA}
\altaffiltext{6}{Astronomical Observatory, Jagiellonian University, ul. Orla 171, 30-244  Krak\'ow, Poland}
\altaffiltext{7}{Nicolaus Copernicus Astronomical Center, Bartycka 18, 00-716 Warsaw, Poland}
\altaffiltext{8}{Institute for Astronomy, Department of Physics, ETH Zurich, CH-8093 Zurich, Switzerland}

\begin{abstract}
Compact radio galaxies with a GHz-peaked spectrum (GPS) and/or compact-symmetric-object (CSO) 
morphology (GPS/CSOs) are increasingly detected in the X-ray domain.
Their radio and X-ray emissions are affected by significant absorption.
However, the locations of the X-ray and radio absorbers are still debated.
We investigated the relationship between the column densities of the total 
(\NHns) and neutral (\NHIns) hydrogen to statistically constrain the picture. 
We compiled a sample of GPS/CSOs including both literature data and new radio data that we acquired with the 
Westerbork Synthesis Radio Telescope for sources whose X-ray emission was either established or under investigation.
In this sample, we compared the X-ray and radio hydrogen column densities, and found that \NH and \NHI display a 
significant positive correlation, with \NHIns$\propto$\NHns$^b$, 
where $b=0.47$ and $b=0.35$, depending on the subsample.
The \NH - \NHI correlation suggests that the X-ray and radio absorbers are either co-spatial or 
different components of a continuous structure.
The correlation displays a large intrinsic spread that we suggest to originate from fluctuations, around a mean value, of the  
ratio between the spin temperature and the covering factor of the radio absorber, $T_{\rm s}/C_{\rm f}$.
\end{abstract}


\section{Introduction}
\label{sec_introduction}

Compact radio galaxies are a class of radio sources whose radio structure is fully contained 
within the host galaxy.
The most compact ones are well sampled by GHz-peaked spectrum (GPS) galaxies and Compact Symmetric Object (CSO) galaxies, two classes of sub-kpc scale 
radio galaxies that largely overlap; the former class is characterized by a spectral turnover observed at frequencies about $0.5-10$ GHz, whereas 
the latter displays a symmetric radio structure whose emission is most often dominated by two mini-lobes.
According to the widely accepted {\it youth scenario}, GPS/CSO galaxies represent the youngest fraction of the radio galaxy population ($<$10$^4$ years):
they would first evolve into the larger, sub-galactic scale compact steep spectrum sources with medium symmetric object morphology (CSS/MSOs), and then possibly further expand beyond their host galaxy, becoming large-scale radio sources \citep{fanti1995,readhead1996,snellen2000}.
However, intermittency of the central engine \citep{reynolds-begelman1997,czerny2009} and possible 
slowing down or disruption of the jet flow \citep[][and references therein]{alexander2000,wagner2012,perucho2016} 
may play an important role in this evolutionary path.
Regardless of whether they are newly born or restarted sources, GPS/CSOs are ideal laboratories to investigate the interplay between the active galactic nucleus (AGN) and the interstellar medium (ISM) in the early phase of the jet activity.

Although still relatively small, the sample of X-ray detected compact radio galaxies is steadily increasing. 
Among them, GPS/CSOs have displayed a very high detection rate ($\sim 100$\%) in a number of X-ray studies during the last decades 
\citep{odea2000,risaliti2003,guainazzi2004,guainazzi2006,vink2006,siemiginowska2008,tengstrand2009,siemiginowska2016};  
one of them was recently detected also in the $\gamma$-ray domain \citep{migliori2016}.
The best angular resolution currently available in the X-ray band ($\sim$1$''$ with {\it Chandra}) is not sufficient to resolve the 
X-ray morphology of most GPS/CSOs. Extended emission has been detected only in two sources, PKS 1345$+$125 
\citep{siemiginowska2008} and PKS 1718$-$649 \citep{siemiginowska2016}; therefore,  
both the nature and the production site of the observed X-rays 
have been mostly investigated through X-ray spectral studies, and are still highly model dependent. 
X-rays in GPS/CSOs have been proposed to be thermal emission from the ISM shocked by the expanding radio lobes \citep{heinz1998,odea2000}, 
thermal Comptonization emission from the disc corona 
\citep{guainazzi2004,vink2006,guainazzi2006,siemiginowska2008,siemiginowska2016,tengstrand2009},
or non-thermal emission of compact lobes produced through inverse-Compton scattering of the local radiation fields
\citep{stawarz2008,ostorero2010,siemiginowska2016}.
All these thermal and non-thermal components are, in fact, likely to contribute to the total X-ray emission of  
compact radio galaxies, but they are difficult to disentangle \citep{siemiginowska2009,siemiginowska2016,tengstrand2009}.

Both radio and X-ray emissions often appear to be affected by significant absorption within the sources, and   
the absorbers may be characterized by complex structures and geometries, as detailed in Section \ref{sec_context}.
The neutral hydrogen column density of the radio absorber, \NHIns, can be estimated from \HI absorption measurements, and the total hydrogen equivalent column density of the X-ray absorber, \NHns, can be derived from X-ray spectral studies. 

The comparisons presented in the literature between \NH and \NHI in GPS/CSOs
indicate that the \NH values are systematically larger than the \NHI values by 1--2 orders of magnitudes \citep[e.g.,][]{vink2006,tengstrand2009}.
Furthermore, a significant positive correlation between \NH and \NHI 
was discovered by \citet{ostorero2009,ostorero2010}: in a sample of 10 GPS/CSOs, they found 
\NH $\propto$ \NHIns$^{\alpha}$, where $\alpha \simeq 1$.
This correlation is indeed expected if the radio and X-ray absorbers are physically connected and the physical and geometrical properties of the absorbers are comparable in different GPS/CSOs. 
Conversely, no correlation is expected if the two absorbers are not physically connected.
Therefore, the possible relationship between \NH and \NHI deserves to be carefully investigated.

Motivated by this finding, and with the aim of investigating the \NH - \NHI relationship for the whole sample of GPS/CSOs known 
to be X-ray emitters, we carried out a program of observations with the Westerbork Synthesis Radio Telescope (WSRT)
aimed at searching for \HI absorption in the GPS/CSOs still lacking an \HI detection.
Preliminary results of this project were presented in \citet{ostorero2016}.

The paper is organized as follows:
in Section \ref{sec_context}, we present the main physical and observational aspects of the X-ray and \HI absorption measurements.
In Section \ref{sec_HIdata}, we present the observations and data analysis of the 
source sample that we observed with the WSRT.
In Section \ref{sec_sample}, we review the source sample that is the subject of our \NH - \NHI investigation.
In Sections \ref{sec_correlation} we present the correlation analysis.
We discuss our results in Section \ref{sec_discussion}, 
and  we draw our conclusions in Section~\ref{sec_conclusions}.

\section{Physical and observational aspects of the absorption}
\label{sec_context}

In the radio band, observations of the spin-flip transition of neutral atomic hydrogen (\HIns) in absorption, at the rest-frame frequency of 1.420 GHz ($\lambda$=21 cm),\footnote{This process is usually referred to as associated absorption.} are a powerful tool to probe the neutral, atomic ISM.
Several \HI absorption surveys revealed that compact radio galaxies 
display a significant excess in the detection rate with respect to extended radio sources \citep{conway1997,morganti2001,
pihlstroem2003,vermeulen2003,gupta2006,chandola2011,curran2013,gereb2014}.

In particular, 
\cite{curran2013} were able to associate this excess with compact sources characterized by projected linear sizes of 0.1--1 kpc (detected with a rate $\gtsim 50$\%, compared to a rate $\ltsim 30$\% for sources with either smaller or larger sizes).
This finding may indicate that the typical cross-section of cold, absorbing gas is 0.1$-$1 kpc, in ``resonance'' with the radio source size.
The detection rate of \HI absorption was also found to be partly affected by the UV luminosity of the AGN: in sources with $L_{UV}> 10^{23}$ W Hz$^{-1}$, mostly extended, a larger fraction of the hydrogen reservoir may be ionized, 
decreasing the likelihood of \HI detection \citep{curran-whiting2010,allison2012}.
The statistics of \HI detections thus seem to suggest that compact radio sources are hosted by 
systems that are, on average, richer in neutral gas than extended sources and that this gas is mostly concentrated in 
structures with typical linear size of 0.1$-$1 kpc,
i.e., larger than the pc-scale dusty tori required by unification schemes \citep{antonucci1993,urry-padovani1995,tadhunter2008}
and recently imaged in nearby Seyfert galaxies \citep[e.g.,][]{jaffe2004,raban2009,tristram2009,tristram2014}. 

Absorbers with a size of 0.1--1 kpc were also shown to best account for the anticorrelation between the peak observed 
optical depth, $\tau_{\rm obs,peak}$, and the projected linear size 
of radio sources \citep{curran2013}: for a given intrinsic optical depth, $\tau$, this anticorrelation arises from 
the proportionality between $\tau_{\rm obs,peak}$ and the fraction of the source covered by the absorber (i.e., the 
covering factor, $C_{\mathrm f}$), and is thus a mere consequence of geometry.
This anticorrelation also drives the anticorrelation between \HI column density, \NHIns, and projected linear size discovered by \cite{pihlstroem2003} for compact radio sources.

However, the actual geometry and dynamical state of the absorbing gas are still a matter of debate.
\HI absorption spectra of compact sources reveal not only a wide range of observed optical depths 
($\tau_{\rm obs,peak}\sim$0.001--0.9), 
but also a remarkable variety of line profiles (Gaussian, multi-peaked, irregular). These
profiles are characterized (i) by widths spanning from less than 
10 km \persec to more than $\sim$1000 km \persec (with typical values of 100-200 km \persec), 
and (ii) by spectral velocities either coincident with the systemic velocity of the galaxy or red/blue-shifted 
up to $\sim$1000 km \persec.  
Prominent red or blue wings spanning several 100 km \persec are also detected in some sources  
\citep[e.g.,][]{vermeulen2003,morganti2003,morganti2004,glowacki2017}.
All this evidence suggests that the kinematics of the absorbing gas can be complex. 
Specifically, as shown by \citet{gereb2014,gereb2015} and by \citet{glowacki2017}, the tendency of compact sources to have broader, deeper, more asymmetric, and more commonly blue-/red-shifted absorption line profiles than extended sources likely reflects the presence of unsettled gas distributions, possibly generated by the interaction of the expanding jets with the circumnuclear medium. 
On the other hand, a fraction of compact sources seems to be depleted of cold atomic gas, as revealed by stacking techniques applied to the spectra of non-detected sources \citep{gereb2014}; the nature of this dichotomy is not clear yet.  

In the limited sample of compact sources with available high angular resolution \HI absorption measurements, 
the \HI is typically detected against one or both radio lobes \citep[][and references therein]{araya2010,morganti2013}, 
although not against the radio core \cite[see, however,][]{peck-taylor2001}. 
On the other hand, the core is often very weak or undetected at frequencies of 5 GHz or higher \citep{taylor1996,araya2010}.
The estimated \HI covering factors, $C_{\mathrm f}$, vary from $\sim$0.2 \citep{morganti2004,morganti2013} 
to $\sim$1 \citep{peck1999}. The general consensus is that the absorber 
has an inhomogeneous or clumpy structure, although the actual distribution of the gas is not known. 
The proposed scenarios include circumnuclear, clumpy tori \citep[e.g.,][]{peck-taylor2001}
and inclined, thin, clumpy discs \citep[e.g.,][]{araya2010} with sizes of $\sim$100 pc,
and/or clouds interacting with the expanding jets and lobes \citep{morganti2004,morganti2013}, 
falling toward the nucleus \citep{conway1999}, or located in a kpc-scale galactic disc \citep{perlman2002}
that might be randomly oriented with respect to the pc-scale torus \citep{curran2008,emonts2012}. 
Evidence of circumnuclear atomic discs with sizes of $\sim$100 pc
and/or infalling clouds was also found in the central regions of extended radio galaxies, including Centaurus A 
\citep{morganti2008} and Cygnus A \citep{struve2010}. 

In the X-ray band, the spectra of GPS/CSOs reveal a significant, although moderate, degree of intrinsic absorption.
\cite{tengstrand2009} found that the mean column density value of their GPS/CSO sample,  
\NH $=3\times 10^{22}$ cm$^{-2}$ ($\sigma_{N_H}\simeq 0.5$ dex), 
is much higher than that of a control sample of extended radio galaxies of the FR-I type 
(\NH $=3.3_{-0.7}^{+2.1}\times 10^{21}$ cm$^{-2}$),
whose core emission does not appear to be obscured by dusty tori \citep{chiaberge1999,donato2004},
and is intermediate between the values of unobscured (\NH $\ltsim 10^{22}$ cm$^{-2}$) and highly obscured 
(\NH $\gtsim 10^{23}$ cm$^{-2}$) FR-II radio galaxies, where the presence of optically thick tori is supported by 
optical and X-ray observations \citep{sambruna1999,chiaberge2000}.
On the other hand, in the CSO sample analyzed by \citet{siemiginowska2016}, 
there appears to be an overabundance ($\sim 60$\%) of sources with mild intrinsic obscuration, \NH $<10^{22}$ cm$^{-2}$.
The mean column density of the full sample is \NH $\simeq (2-4) \times 10^{21}$ cm$^{-2}$ ($\sigma_{N_H}\simeq 0.3$ dex), 
depending on whether only detections or both detections and upper limits are taken into account;
these values are consistent with the hydrogen column densities of FR-I and unobscured FR-II radio galaxies that we mentioned above. 

The geometry and scales of the X-ray obscuring circumnuclear structures in AGN are still debated: 
the ratio between Type-2 and Type-1 AGN requires a geometrically thick torus,
that is modeled in different ways, e.g. as a pc-scale, dusty donut \citep[e.g.,][]{krolik1988}, or as a   
dust-free, sub-pc scale, clumpy outflow closely related to the broad-line emission region 
\citep{risaliti2007,risaliti2010,risaliti2011,nenkova2008}. 
Whether or not the X-ray obscuration of GPS/CSOs fits any of these scenarios is not clear, 
and any relationship between the radio and X-ray absorbers may help to clarify the picture.

As mentioned in Section \ref{sec_introduction}, the neutral hydrogen column densities, \NHIns, 
appear to be systematically lower than the total hydrogen equivalent column densities, \NHns, 
by 1--2 orders of magnitudes, in GPS/CSOs for which both X-ray and spatially unresolved \HI spectra are available
\citep[e.g.,][]{vink2006,tengstrand2009}. 
This discrepancy may indicate that the X-ray and radio measurements trace different absorbers, in agreement 
with a scenario where the X-rays originate in the accretion-disc corona, and are consequently more affected by obscuration 
than the radio waves with $\lambda=21$ cm produced farther from the AGN \citep[e.g.,][]{vink2006,tengstrand2009}.
However, care should be taken when comparing the radio and X-ray hydrogen column densities. 

First, the estimate of \NHI depends on the ratio between the spin temperature of the absorbing 
gas and the covering factor, $T_{\mathrm s}/C_{\mathrm f}$ \citep[e.g.][]{wolfe1975,odea1994,gallimore1999}, a parameter that is often poorly constrained; common assumptions are $T_{\mathrm s}=100$ K and $C_{\mathrm f}=1$, suitable for a cold
gas cloud in thermal equilibrium, and thus with the spin temperature equal to the kinetic temperature ($T_{\mathrm s}=T_{\mathrm k}$),
that fully covers the radio source.
Secondly, the \HI absorption measurements trace the content of the absorbing system in terms of neutral hydrogen (\NHIns),
whereas the X-ray spectral measurements enable to constrain, for a given X-ray emission model, the content of total (i.e., neutral, 
molecular, and ionized) hydrogen (\NH$=$\NHI$+$\NHns$_2$$+$\NHIIns) in an absorber with a given elemental abundance 
\citep[e.g.,][]{wilms2000}; full coverage of the X-ray source by the absorber is also typically assumed.
A difference between \NHI and \NH is thus expected in a single absorbing system composed of gas that is not fully atomic and neutral;
this difference is ultimately set by the temperature of the gas \citep[e.g.,][]{maloney1996}. 

When the absorber is cold, and the assumption $T_{\mathrm k} = T_{\mathrm s}=100$ K is reasonable for the \HI gas, the abundance of 
molecular hydrogen may be high \citep[e.g.,][]{maloney1996} and may partly account for the \NH - \NHI discrepancy. 
On the other hand, when the neutral hydrogen is warmer than typically assumed, with $T_{\mathrm s}\simeq$ few $10^3$ K, as expected in the circumnuclear AGN environment \citep{bahcall1969,maloney1996,liszt2001},  the \NHI estimated from the observed optical depth $\tau_{obs}$ increases accordingly, and may become comparable to the \NH estimates.  

The assumption of a partial coverage of the source by the absorber ($C_{\mathrm f}<1$) affects both the \NHI and the \NH estimates, 
but the details of  this effect are still to be investigated.

Despite the above uncertainties, which affect the magnitude of the \NH - \NHI offset, the existence of a correlation 
between \NH and \NHI \citep{ostorero2010,ostorero2016}, which we confirm below, clearly points toward a physical 
connection between the X-ray and radio absorbers.

\section{\HI observations and data analysis}
\label{sec_HIdata}

\subsection{Target sample}
\label{subsec_targets}
A summary of our observing program with the WSRT is reported in Table \ref{tab_wsrt}. 
We searched for \HI absorption in a sample of 12 GPS/CSOs,
hereafter referred to as the {\it target sample},
drawn from a larger sample of X-ray emitting GPS/CSOs described in Section \ref{sec_sample}.
The target sources are listed in Table \ref{tab_wsrt}, Column 1 
(and are also marked with an asterisk in the last column of Table \ref{tab_NHI}). 
Their optical redshifts are reported in Column 2 of the same Table.

Four out of twelve targets (i.e., 0035$+$227, 0026$+$346, 2008$-$068, and 2128$+$048) were not searched for \HI absorption prior to our study. 
In the remaining eight targets, \HI absorption was not detected in previous observations, 
and upper limits to the \HI optical depth could be estimated for five of them 
(see Table \ref{tab_NHI}). We observed these eight sources again to either improve the available upper limits or attempt 
the first estimate of their \HI optical depths.

\subsection{Observations}
\label{subsec_observations}

The setup of the observations is summarized in Table \ref{tab_wsrt}, Columns 3--6.

The observations were carried out with the WSRT in five observing runs from  2008 to 2011.
The target sources were observed either with the UHF-high-band receiver (appropriate when $z \gtsim 0.2$) 
or with the L-band receiver (appropriate when $z<0.2$), in dual orthogonal polarization mode.
Each target was observed for an exposure time of 3.5$-$12 hours.
The observing band was either 10 or 20 MHz wide,
with 1024 spectral channels, and was centered at the 
frequency $\nu_{\rm obs}$ where the \HI absorption line is expected to occur, based on the spectroscopical optical redshift.
Compared to the \HI survey of compact sources by \citet{vermeulen2003}, where a 10 MHz 
wide band and 128 spectral channels were available, 
our observations could benefit from a larger ratio between number of spectral channels and observing 
bandwidth; this improvement, together with the longer exposure times, had the potential to enable 
a more effective separation of narrow \HI absorption features from radio frequency interferences (RFI).
However, in many cases, we were forced to adopt a 10 MHz wide band in order to minimize the in-band effect of nearby 
RFI.\footnote{The frequencies corresponding to the redshifts of these sources are close to the GSM band and bands allocated to TV broadcast.}
The 10 MHz bandwidth enabled us to cover a velocity range about the velocity centroid of approximately 
$\pm$ 1200 km \persec at $z\simeq 0.1$, increasing to approximately $\pm$2000 km \persec at $z\simeq 1$.
When we could use a 20 MHz wide band, the above velocity ranges increased to approximately $\pm$ 2300 km \persec 
and $\pm$ 4200 km \persec, respectively.

A primary calibrator (either 3C~48, 3C~147 or 3C~286) was observed before and after each target pointing, and used as a flux and bandpass calibrator.

\subsection{Data analysis}
\label{subsec_dataanalysis}

The integration time of our observations was typically of only a few hours (see Table \ref{tab_wsrt}). 
For an east-west array like the WSRT, this implies that the synthesized beam of the data cubes is very elongated. However, our targets are all 
unresolved by the WSRT; therefore, this is not an issue for the results presented in this work. 

The data were calibrated and reduced using the MIRIAD package \citep{sault1995}. The continuum subtraction was done by performing a linear fit 
of the spectrum through the line-free channels of each visibility record, and subtracting the fitting function from all the frequency channels 
(``UVLIN''). The spectral-line cube was obtained by averaging a few (typically three or four) channels together and adopting uniform weighting. 
The data were Hanning smoothed to suppress the Gibbs ripples.
The final velocity resolution of the data cubes, together with the 3$\sigma$ rms noise, are listed in Table \ref{tab_wsrt}, Columns 7 and 8.

For all but one target source, we used the UHF-high-band receiver: this receiver, unlike the L-band receiver,
is uncooled. Therefore, despite the technical improvements described in Section \ref{subsec_observations},
our observations were affected by a relatively high noise level.

\subsection{HI results}
\label{subsec_HIresults}

We detected \HI absorption in 2 out of 12 targets, 0035$+$227 and 0941$-$080.
For seven targets, we could estimate upper limits to the optical depth of a putative \HI absorption line.
However, in three cases, the quality of our observations turned out to be lower than the quality of the observations 
previously performed by \citet{vermeulen2003}. This appears to be due to degradation of the band caused by extra RFI. 
In these three cases, we used the constraints on the optical depth derived by \citet{vermeulen2003} for our correlation 
analysis (see Section \ref{sec_correlation}).
For the remaining three targets, located at redshift $z=0.517-0.547$, 
the RFI were too strong to obtain any useful data, despite the above technical improvements. 

For the sources in which we detected \HI absorption, the peak optical depth, $\tau_{\rm obs,peak}$, the line width, $\Delta V$ 
(corresponding to the full width at half maximum), and the peak velocity, $v_{\rm obs,peak}$, were determined through 
Gaussian fitting of the absorption profiles.
For the sources in which no \HI absorption was detected, $3\sigma$ upper limits to $\tau_{\rm peak}$ were estimated
from the continuum flux density and the rms noise level of the spectra.
The estimates of $\tau_{\rm obs,peak}$, $\Delta V$, and $v_{\rm peak}$ are reported in Table \ref{tab_wsrt}, 
Columns 11, 12, and 13, respectively. 
The flux density of the continuum and, for the detections only, its difference from the flux density of the line peak 
are given in Columns 9 and 10, respectively.

The atomic hydrogen column density along the line of sight (in units of \percmq) is related to the velocity-integrated optical 
depth of the \HI absorption profile through the following relationship \citep[e.g.,][]{wolfe1975,odea1994,curran2013}:
\begin{equation}
\label{eq_NHI}
N_{\rm HI} =1.823\times 10^{18}\, T_{\mathrm s} \int{\tau(v) \,{\mathrm d}v} , 
\end{equation}
where $T_{\mathrm s}$ is the spin temperature in K, $v$ is the velocity in units of km \persec, and the optical depth is given by 
\begin{equation}
\label{eq_tau}
\tau(v) = - \ln \left[1-\frac{\tau_{\rm obs}(v)}{C_{\mathrm f}}\right].
\end{equation}
Here, the observed optical depth, $\tau_{\rm obs}(v)$, is the ratio between the spectral line depth 
in a given velocity channel 
and the continuum flux density of the background radio source: 
$\tau_{\rm obs}(v)\equiv \Delta S(v)/S_{\rm cont}(v) = (S(v)-S_{\rm cont}(v))/S_{\rm cont}(v)$.

In the optically thin regime (i.e., for $\tau \ltsim 0.3$), 
the observed optical depth of the line is related to the actual optical depth through 
$\tau(v)\approx \tau_{\rm obs}(v)/C_{\mathrm f}$, and Equation (\ref{eq_NHI}) can be approximated by
\begin{equation}
\label{eq_NHI_approx}
N_{\rm HI} \approx 1.823\times 10^{18}\, (T_{\mathrm s}/C_{\mathrm f}) \int{\tau_{\rm obs}(v)\, {\rm d}v} .
\end{equation}
Therefore, the \HI column density can be derived from the integrated observed optical depth, 
$\tau_{\rm obs}\equiv \int{\tau_{\rm obs}(v)\, {\rm d}v}$, once 
the ratio $T_{\mathrm s}/C_{\mathrm f}$ is known.

We computed the \HI column densities (in \percmq) of the detected sources from Equation (\ref{eq_NHI_approx}),
with $\tau_{\rm obs}(v)$ replaced by its Gaussian fitting function.
For consistency with the literature,
we estimated \NHI by assuming that the absorber fully covers the 
radio source ($C_{\mathrm f}$=1), and that the spin temperature of the absorbing gas 
is $T_{\mathrm s}$=100 K. These assumptions yield \NHI values that likely represent lower limits to the actual column densities 
(see Section \ref{sec_context}). 

When we  did not detect any \HI line, we used the following equation to estimate $3\sigma$ upper limits to \NHI from the $3\sigma$ upper limits to the observed optical depth, 
$\tau_{\rm obs,3\sigma}$:
\begin{equation}
\label{eq_NHIobsUL}
N_{\rm HI,3\sigma} = 1.823\times 10^{18}\, (T_{\mathrm s}/C_{\mathrm f})\, \tau_{\rm obs,3\sigma}\,\Delta V , 
\end{equation}
under the assumption that the putative absorption line has $\Delta V=100$ km \persec \citep{vermeulen2003}.
Our estimates of \NHI are reported in Table \ref{tab_wsrt}, Column 14.

\subsubsection{HI absorption in 0035$+$227}
\label{subsubsec_0035HI}

\begin{figure}[!h]
\includegraphics[scale=0.4]{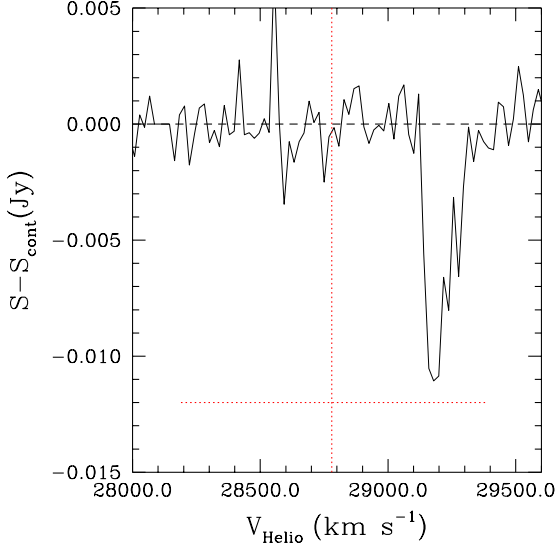}
\caption{Spectrum of the CSO 0035$+$227, with velocities displayed in the optical heliocentric convention.
The systemic velocity of the optical galaxy is $v_{\rm sys}=(28,780\pm 600)$ km \persec: $v_{\rm sys}$ is 
marked with a vertical, dotted line; the horizontal, dotted line shows its 1$\sigma$ uncertainty.
A complex absorption line, with $\tau_{\rm obs,peak}=0.0181\pm 0.0014$ and $\Delta V=(114\pm 10)$ km \persec, 
is detected about $v_{\rm sys}$.}
\label{fig_0035HI}
\end{figure}

Our spectrum of source 0035$+$227 is displayed in Figure \ref{fig_0035HI}.
We detected one absorption feature with a complex profile. A Gaussian fit to the optical depth velocity profile
yields a peak optical depth $\tau_{\rm obs,peak}=0.0181\pm 0.0014$, a line width $\Delta V=(114\pm 10)$ km \persec, and  
a peak velocity $v_{\rm peak}= (29197 \pm 4)$ km \persec.
The systemic velocity of the host galaxy, derived from its heliocentric redshift, $z=0.096\pm0.002$, 
estimated by \citet{marcha1996} from an optical spectrum of the source, is $v_{\rm sys}=(28780\pm 600)$ km \persec.
The peak velocity of the absorption line is consistent with the systemic velocity of the galaxy at 1$\sigma$ 
confidence level.
We derived an \HI column density \NHI=$(3.99\pm 0.49) \times 10^{20}$ \percmq  for the gas responsible for the absorption line, 
under the assumption that $C_{\rm f}=1$ and $T_{\rm s}=100$ K.

\clearpage
\LongTables 
\begin{landscape}
\begin{deluxetable}{cccccccccccccc}
\tabletypesize{\scriptsize}
\tablecaption{WSRT Observations: Source Sample, Observation Details, Results of the Data Analysis, and Estimates of the \HI Column Density.\label{tab_wsrt}}
\tablewidth{0pt}
\tablehead{
\colhead{Source Name} &
\colhead{$z_{\mathrm {opt}}$\tnm{a}} &
\colhead{Receiver} &
\colhead{$t_{\mathrm {exp}}$} &
\colhead{Bandwidth} &
\colhead{$\nu_{\mathrm {obs}}$} &
\colhead{Spectral Res.} &
\colhead{rms} &
\colhead{$S_{\mathrm {cont}}$} &
\colhead{$S_{\mathrm {cont}} - S_{\mathrm {HI,peak}}$} & 
\colhead{$\tau_{\mathrm {obs,peak}}$} &
\colhead{$\Delta V$} &
\colhead{$v_{\rm peak}$} & 
\colhead{\NHI} \\
\colhead{B1950} &
\colhead{~~} &
\colhead{~~} &
\colhead{(h)} &
\colhead{(MHz)} &
\colhead{(MHz)} &
\colhead{(km \persec)} &
\colhead{(mJy/b)} &
\colhead{(Jy)} &
\colhead{(mJy)} &
\colhead{($10^{-2}$)} &
\colhead{(km \persec)} &
\colhead{(km \persec)} &
\colhead{\big($10^{20}\frac{T_{\mathrm s}}{100 \mathrm{\,K}}$ \percmq \big)} \\
\colhead{(1)} &
\colhead{(2)} &
\colhead{(3)} &
\colhead{(4)} &
\colhead{(5)} &
\colhead{(6)} &
\colhead{(7)} &
\colhead{(8)} &
\colhead{(9)} &
\colhead{(10)} &
\colhead{(11)} &
\colhead{(12)} &
\colhead{(13)} &
\colhead{(14)} \\
}
\startdata
0019$-$000 & 0.305  & UHF-high & 5 & 10 & 1088    & 16      & 7.8     & 2.8     & \nodata & $<0.84$ & \nodata & \nodata  & $<1.5$\tnm{b,c}\\  
0026$+$346 & 0.517 & UHF-high & 4   & 10 &  936.3   & \nodata & \nodata & \nodata & \nodata & \nodata & \nodata & \nodata & \nodata \\
0035$+$227 & 0.096$\pm$0.002 & L-band & 4  & 20 &  1296        & 20  & 1.5     & 0.583    & 11     & $1.81\pm 0.14$   & $114\pm10$ & $29,197 \pm 4$ & $3.99\pm0.49$ \\
0710$+$439 & 0.518 & UHF-high & 4   & 10 &  935.7   & \nodata & \nodata & \nodata & \nodata & \nodata & \nodata & \nodata & \nodata \\
0941$-$080 & 0.2281$\pm$0.0013\tnm{d} & UHF-high & 5 & 10 &  1156.7 & 8       & 1.7     & 2.58    & 7   & $0.22\pm0.02$ & $215\pm 18$ & $68,156\pm8$ & $0.91\pm0.10$ \\ 
1031$+$567 & 0.459\tnm{e} & UHF-high & 3.5 & 20    &  973.6   & 34      & 6.6     & 1.99    & \nodata & $<0.99$ & \nodata & \nodata & $<1.8$\tnm{b,c} \\
1117$+$146 & 0.362 & UHF-high & 5  & 10 &  1043    & 16      & 5.6     & 2.7     & \nodata & $<0.62$ & \nodata & \nodata  & $<1.1$\tnm{b,c}\\
1607$+$268 & 0.473 & UHF-high & 5   & 10 &  964     & 16      & 14      & 4.88    & \nodata & $<0.86$ & \nodata & \nodata & $<1.6$\tnm{b,c}\\
1843+356   & 0.764 & UHF-high & 12 & 20 &  805.2    & 16      & 5.3     & 0.248   & \nodata & $<6.41$ & \nodata & \nodata & $<11.7$\tnm{b,c} \\
2008$-$068 & 0.547 & UHF-high & 5 & 10 &  918.2    & \nodata & \nodata & \nodata & \nodata & \nodata & \nodata & \nodata & \nodata  \\
2021$+$614 & 0.227 & UHF-high & 4 & 20 &  1157.6    & 16      & 0.98    & 1.99    & \nodata & $<0.15$ & \nodata & \nodata & $<0.27$\tnm{b,c}\\
2128$+$048 & 0.99  & UHF-high & 8 & 10 &  713       & 8       & 17      & 4.46    & \nodata & $<1.14$ & \nodata & \nodata & $<2.08$\tnm{b,d}\\
\enddata
\tablenotetext{a}{Redshift uncertainties are reported only for the sources in which \HI absorption was detected, for comparison 
              with the \HI absorption line peak velocity.}
\tablenotetext{b}{3$\sigma$ upper limit.}  
\tablenotetext{c}{From results on $\tau_{\rm obs,3\sigma}$, using the relation: 
                  \NHIns$_{,3\sigma}= 1.823\times10^{18}\times T_{\mathrm{s}} \times \tau_{\rm obs,3\sigma} \times \Delta V$, under the 
                  assumptions  $T_{\mathrm{s}}$=100 K, $\Delta V = 100$ km \persec, and $C_{\mathrm f}=1$.}
\tablenotetext{d}{W.H. de Vries, private communication.} 
\tablenotetext{e}{In a previous paper \citep{ostorero2016}, we used the redshift value $z=0.45$ \citep{pearson1988} for this source. 
                  This redshift is incorrect, and was here replaced by the more appropriate value $z=0.4590\pm 0.0001$ \citep{dunlop1989}.}
\end{deluxetable}
\clearpage
\end{landscape}

\subsubsection{HI absorption in 0941$-$080}
\label{subsubsec_0941HI}

\begin{figure}[!h]
\includegraphics[scale=0.4]{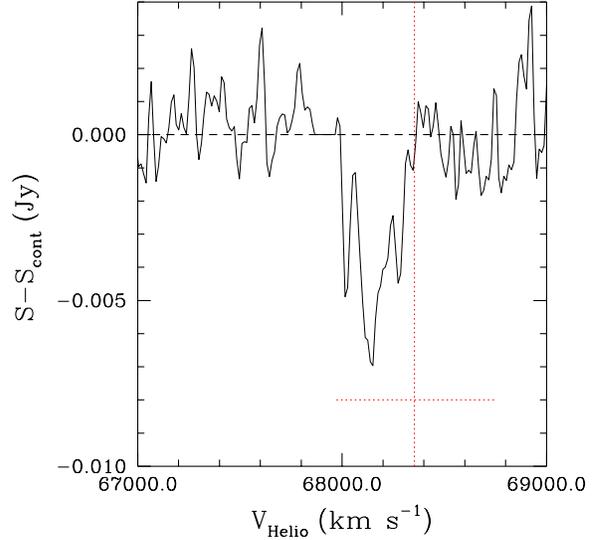}
\caption{Spectrum of the GPS/CSO galaxy 0941$-$080, with velocities displayed in the optical heliocentric convention.
The systemic velocity of the optical galaxy is $v_{\rm sys}=(68,353\pm 390)$ km \persec:  
$v_{\rm sys}$ is marked with a vertical, dotted line; the horizontal, dotted line shows its 1$\sigma$ uncertainty.
A broad, multi-peaked absorption feature with $\tau_{\rm obs,peak}=0.0022 \pm 0.0002$ and $\Delta V=(215 \pm 18)$ km \persec is 
detected about $v_{\rm sys}$.}
\label{fig_0941HI}
\end{figure}

In the spectrum of 0941$-$080, shown in Figure \ref{fig_0941HI}, we detected a broad, multi-peaked absorption 
feature.
A Gaussian fit to the optical depth velocity profile yields a peak optical depth $\tau_{\rm obs,peak}=0.0022 \pm 0.0002$, a line width 
$\Delta V=(215\pm 18)$ km \persec, and a peak velocity $v_{\rm peak}= (68156 \pm 8)$ km \persec.
The systemic velocity of the host galaxy, derived from the heliocentric redshift estimated from the optical 
spectrum of the source, $z=0.2281 \pm 0.0013$ \citep[][de Vries, private communication]{devries2000}, 
is $v_{\rm sys}=(68353\pm 390)$ km \persec. 
The peak velocity of the absorption line is consistent with the systemic velocity of the galaxy at 1$\sigma$ 
confidence level.
For the gas responsible for this associated absorption, we derived an \HI column density 
\NHI=$(9.1\pm 1.0)\times 10^{19}$ \percmq, under the assumption that $C_{\rm f}=1$ and $T_{\rm s}=100$ K. 
This detection of \HI absorption in the source is the first ever, and it was enabled by the improved observation setup
(see Section \ref{subsec_observations}).
Previous observations \citep{vermeulen2003} could only set upper limits to the optical depth of a putative absorption 
line (see Table \ref{tab_NHI}). 
The optical depth that we measured is consistent with the previously estimated upper limits.

\section{\NH - \NHI sample definition} 
\label{sec_sample}

By using our own data and those from the literature, we compiled the largest sample of GPS/CSOs that were targets of both X-ray and \HIns-absorption observations.
The GPS/CSOs observed in the X-rays were all detected in this band, and they constitute a subsample of the GPS/CSOs observed in the \HI band: therefore, the sample that we compiled consists of all the X-ray detected GPS/CSOs that were also observed in the \HI band, i.e. 27 compact radio galaxies.
This sample includes sources that were investigated in the X-ray band, either individually or in small samples, 
by different authors and with different instruments; it can be considered as the merger of two, partly overlapping, subsamples of sources that were selected for X-ray investigations with different criteria. 
The first subsample is a flux- and volume-limited sample of 17 GPS galaxies, with $F_{5\,\mathrm GHz} > 1$ Jy and $z<1$, extracted 
by \citet{tengstrand2009} and \citet{guainazzi2006} from the complete sample of GPS sources compiled by \citet{stanghellini1998}; these GPS galaxies are located at redshift $z=0.0773-0.99$, and have radio luminosities at 
$\nu=1.4$ GHz spanning the range $L_{\mathrm {1.4\,GHz}}= 10^{25}-10^{28.4}$~W \perhz; 16 of them were morphologically classified as CSOs. 
The second subsample is a sample of 16 CSOs with known redshifts and estimated kinematic ages, compiled by \citet{siemiginowska2016}; these CSOs are located at redshift $z=0.0142-0.764$,
and have radio luminosities at $\nu=1.4$ GHz spanning the range $L_{\mathrm {1.4\,GHz}}= 10^{24}-10^{27.6}$ W \perhz; 15 of them 
were spectroscopically classified as GPS sources. The two subsamples have six sources in common.
Overall, the 27 sources of the full sample are located at redshift $z=0.0142-0.99$, and they have moderate to high 
radio luminosities, $L_{\mathrm {1.4\,GHz}}= 10^{24}-10^{28.4}$ W \perhz.\footnote{Only 3 out of 27 sources have $L_{\mathrm {1.4\,GHz}}<10^{25}$ W \perhz: 1245$+$676, 1509$+$054, and 1718$-$649.}
Even though the sample is not complete and well-defined in terms of flux limit and volume, it is the largest sample of 
GPS/CSOs available to date for which both X-ray and \HI observations were carried out.
The main properties of the sources of this sample, including all the available \HI and X-ray column density estimates, 
are summarized in Tables \ref{tab_NHI} and \ref{tab_NH}.

In order to perform the \NH - \NHI correlation analysis, from the above sample of 27 GPS/CSOs we extracted the subsample of sources for which an estimate of both \NHI and \NH was available. 
By ``estimate'' we mean either a value or an upper/lower limit.
The three sources 0026$+$346, 0710$+$439, and 2008$-$068 have no \NHI estimate (see Sections \ref{subsec_NHI} and Table \ref{tab_NHI}),
and the two sources 0116$+$319 and 1245$+$676 have no \NH estimate (see  Sections \ref{subsec_NH} and Table \ref{tab_NH}). 
Therefore, we were left with a subsample of 22 sources, that we refer to  hereafter as the {\it correlation sample}.

The correlation sample spans the redshift range $z=0.0142-0.99$ and the 1.4 GHz luminosity range $L_{\mathrm {1.4\,GHz}}= 10^{24.2}-10^{28.4}$ W \perhz.
Table \ref{tab_NHI_NH} lists the sources of the correlation sample (Column 1), their optical redshifts (Column 2), their radio spectral and morphological classification (Column 3), the estimates of \NHI (Column 4) and \NH (Column 5) that we used for the correlation analysis, the type of (\NHns,\NHIns) pair (Column 6), and the sample to which we associated each (\NHns,\NHIns) pair for the correlation analysis (Column 7), as described in Section \ref{subsec_NH_NHI}.
More details on the quantities reported in Table \ref{tab_NHI_NH} and the criteria that we applied to select the column density estimates given in this table among all the available column density estimates, can be found in the Appendix.

\begin{deluxetable*}{lcccccc}
\tabletypesize{\scriptsize}
\tablecaption{Correlation Sample: Column Density Data Set, Pair Types, and Analysis Samples. \label{tab_NHI_NH}}
\tablewidth{0pt}
\tablehead{
\colhead{Source Name} &
\colhead{$z_{\mathrm {opt}}$} &
\colhead{GPS/CSO} &
\colhead{\NHI} &
\colhead{\NH} &
\colhead{Pair Type} &
\colhead{Sample\tnm{a}} \\
\colhead{B1950} &
\colhead{~~} &
\colhead{~~} &
\colhead{\big($10^{20}\frac{T_{\mathrm s}}{100 \mathrm{\,K}}$ \percmq \big)} &
\colhead{\big($10^{22}$ \percmq \big)} &
~~~    &
~~~    \\
\colhead{(1)} &
\colhead{(2)} &
\colhead{(3)} &
\colhead{(4)} &
\colhead{(5)} &
\colhead{(6)} &
\colhead{(7)} \\
}
\startdata
0019$-$000 & 
0.305 		 & 
GPS   		 & 
$<1.5$        &
$<100$           &
{\it UU}       	 &
$E'$, $E''$    	\\
0035$+$227 & 
$0.096\pm0.002$  & 
CSO      	 & 
$3.99\pm0.49$ 	 & 
$1.4^{+0.8}_{-0.6}$ & 
{\it VV} & 
$E'$, $E''$	\\
0108$+$388 	& 
0.66847 	& 
GPS,CSO  	& 
$80.5$ 		& 
$47.5 ^{+12}_{-12}$\tnm{~b}  & 
{\it VV}       	 &
$E'$       \\ 
~~~             &  
~~~     	& 
~~~     	&  
$80.5$ 		& 
$>90$\tnm{~c}    &   
{\it VL}       	&
--               \\ 
0428$+$205 	& 
0.219  		& 
GPS,CSO    	&  
3.45\tnm{~d}     & 
$<0.69$         &
{\it VU}       	&
$E'$, $E''$             \\ 
0500$+$019 	& 
0.58457 	& 
GPS,CSO   	& 
$6.2$  		& 
$0.5^{+0.18}_{-0.16}$ &
{\it VV}       	&
$E'$, $E''$    	\\ 
0941$-$080  	& 
0.2281$\pm$0.0013 & 
GPS,CSO   	& 
$0.91\pm0.10$  	& 
$<1.26$		&
{\it VU}       	&
$E'$, $E''$            	\\ 
1031$+$567 	& 
0.459   	& 
GPS,CSO  	& 
$<1.26$ 	& 
$0.50\pm0.18$   & 
{\it VU}       	&
$E'$, $E''$      	\\ 
1117$+$146 	& 
0.362  		& 
GPS,CSO 	&  
$<0.63$		& 
$<0.16$ 	&
{\it UU}       	&
$E'$, $E''$        	\\ 
1323$+$321 	& 
0.370  		& 
GPS,CSO 	& 
$0.71$ 		& 
$0.12^{+0.06}_{-0.05}$ & 
{\it VV}       	&
$E'$, $E''$  	\\
1345$+$125 	& 
0.12174 	& 
GPS,CSO 	&  
$6.2$  		& 
$4.8\pm0.4$     & 
{\it VV}       	&
$E'$, $E''$     	\\ 
~~~~~~~         &
~~~~~~~         &
~~~~~~~         &
$6.2$  		& 
$2.543^{+0.636}_{-0.580}$ &
{\it VV}       	&
$E'$, $E''$            \\
1358$+$624 	& 
0.431  		& 
GPS,CSO  	& 
$1.88$ 		& 
$2.9^{+2}_{-1}$ & 
{\it VV}       	&
$E'$, $E''$     	\\ 
1404$+$286 & 
0.07658 	& 
GPS,CSO 	&  
$8.0$  		& 
$0.13^{+0.12}_{-0.10}$  &
{\it VV}       	&
$E'$	\\ 
~~~~~~~         &
~~~~~~~         &
~~~~~~~         &
$8.0$  		& 
$>90$\tnm{~c}	&
{\it VL}       	&
--         \\
1509$+$054 	& 
0.084  		& 
GPS,CSO 	&   
$<3.64$ 	&
$<0.23$  	&
{\it UU}       	&
$E'$, $E''$           	\\ 
1607$+$268 	& 
0.473   	& 
GPS,CSO 	&  
$<1.6$	& 
$<0.18$         &
{\it UU}       	&
$E'$, $E''$            	\\ 
1718$-$649  	&  
0.0142  	& 
GPS,CSO 	&  
1.477\tnm{~d}    &
$0.08\pm 0.07$  & 
{\it VV}       	&
$E'$, $E''$ 		 \\  
1843+356  	& 
0.764   	& 
GPS,CSO 	& 
$<10.4$ 	& 
$0.8^{+0.9}_{-0.7}$ & 
{\it VU}   & 
$E'$, $E''$	        \\ 
1934$-$638 	& 
0.18129 	& 
GPS,CSO 	&  
$0.06$ 		&
$0.08^{+0.07}_{-0.06}$        &
{\it VV}       	&
$E'$           \\
~~~~~~~         &
~~~~~~~         &
~~~~~~~         &
$0.06$ 		&
$>250$\tnm{~c} & 
{\it VL}       	&
--              \\ 
1943$+$546 	& 
0.263  		& 
GPS,CSO 	&  
$4.91$ 		& 
$1.1\pm0.7$  & 
{\it VV}     &
$E'$, $E''$       \\ 
1946$+$708 	& 
0.101   	& 
GPS,CSO 	&  
$31.6$ 		& 
$1.7^{+0.5}_{-0.4}$        &
{\it VV}       	&
$E'$              \\
~~~~~~~         &
~~~~~~~         &
~~~~~~~         &
$31.6$ 		& 
$>280$\tnm{~c} 	& 
{\it VL}       	&
--              	\\ 
2021$+$614 	& 
0.227   	& 
GPS,CSO 	& 
$<0.27$     	& 
$<1.02$ 	& 
{\it UU}       	&
$E'$, $E''$           	\\
2128$+$048 	& 
0.99   		& 
GPS,CSO 	& 
$<2.08$ 	& 
$<1.9$          & 
{\it UU}       	&
$E'$, $E''$           	\\ 
2352$+$495 	& 
0.2379 		& 
GPS,CSO 	& 
2.84\tnm{~d} & 
$4^{+7}_{-3}$   & 
{\it VV}       	&
$E'$, $E''$              \\ 
\enddata
\tablenotetext{a}{The en-dash indicates that the corresponding pair was not included in any sample because it comprises a lower limit (see Section \ref{subsec_NH_NHI} for details).}
\tablenotetext{b}{This value of \NH was estimated as the mean of the $3\sigma$ lower limit \NH$>5\times10^{22}$\percmq and the physical upper bound of the Compton-thin 
                  \NH range, i.e. \NH$>9\times 10^{23}$\percmq (see Appendix \ref{subsec_NHapp} and Table \ref{tab_NH} for details).} 
\tablenotetext{c}{This value of \NH corresponds to the assumption that the absorber is Compton-thick (see Appendix \ref{subsec_NHapp}
                        and Table \ref{tab_NH} for details).}
\tablenotetext{d}{Total \NHIns, estimated as the sum of the \NHI values derived from the two absorption lines detected in the spectrum (see Appendix \ref{subsec_NHIapp}, 
Table \ref{tab_NHI}).}
\end{deluxetable*}

\subsection{\NHI estimates}
\label{subsec_NHI}

Detections of \HI absorption features, and their corresponding \NHI values, were available for 14 out of the 
22 sources of the correlation sample; \NHI upper limits could be estimated for the remaining eight sources.

We only used \NHI estimates derived from low angular resolution measurements, i.e. from measurements that were 
not able to spatially resolve the source; all our upper limits to \NHI are 3$\sigma$ limits.
As mentioned in Sections \ref{sec_context} and \ref{subsec_HIresults}, the \NHI estimates depend on the value assumed 
for the ratio $T_{\mathrm s}/C_{\mathrm f}$.
When an \NHI value drawn from the literature was estimated by the authors by assuming $T_{\mathrm s}/C_{\mathrm f}>100$ K 
(i.e., $T_{\mathrm s}>100$ K and $C_{\mathrm f}=1$), we rescaled it to an \NHI value computed with 
$T_{\mathrm s}/C_{\mathrm f}=100$~K.
For the 12 sources with more than one \NHI estimate, we chose the most suitable estimate, according to the criteria described in Appendix \ref{subsec_NHIapp}.
The selected \NHI estimates are summarized in Table \ref{tab_NHI_NH}, Column 4; they are also listed with the corresponding references in Table \ref{tab_NHI}; we thoroughly comment on these data, as well as on on the properties of the full \NHI data set, in Appendix \ref{subsec_NHIapp}.

\subsection{\NH estimates}
\label{subsec_NH}
As anticipated in Section \ref{sec_context}, the estimate of the column density of the X-ray absorbing gas located at the redshift of the source (i.e., the local absorber) depends on the X-ray emission model adopted to interpret the X-ray spectrum of the source, as well as on the assumed photoionization cross-section of the ISM.
For a given abundance of chemical elements in the absorbing gas, the X-ray spectral modeling 
enables to estimate the equivalent hydrogen column density of the local absorber, \NH, i.e., the column density of hydrogen atoms, molecules, and ions toward the source, located at the source redshift (see Appendix \ref{subsec_NHapp} for details). 

In the correlation sample, detections of intrinsic X-ray absorption, and corresponding values of \NHns, were available for nine out of 22 sources; upper limits to \NH were available for eight sources. 
For the remaining five sources, multiple exposures and ambiguity in the spectral modeling prevented us from selecting a 
single robust \NH value.
In particular, for one source, we selected two significantly different \NH values corresponding to different observational epochs.
For four of them, the ambiguity between a Compton-thin and a Compton-thick absorber in the model 
adopted for the interpretation of the X-ray spectrum of the source led to the availability of \NH estimates 
lower and greater than $\simeq 10^{24}$ \percmq, respectively; we considered both scenarios as plausible, 
and associated with each of these four sources either an \NH value (Compton-thin scenario) or a lower limit 
to \NH (Compton-thick scenario).

The \NH estimates that we used for our correlation analysis are summarized in Table \ref{tab_NHI_NH}, Column 5;
they are also listed with the corresponding references in Table \ref{tab_NH}; we thoroughly comment on these data, as well as on the properties of the full \NH data set, in Appendix \ref{subsec_NHapp}.

\subsection{\NH - \NHI sample}
\label{subsec_NH_NHI}

The correlation sample consists of 22 sources associated with (\NHns,\NHIns) pairs of estimates 
of four different types: pairs of {\it values} (hereafter referred to as {\it VV} pairs), pairs including a value and an upper limit ({\it VU} pairs), pairs of upper limits ({\it UU} pairs), and pairs including a value and a lower limit ({\it VL} pairs).
Specifically, the sample includes four sources unambiguously associated with {\it VU} pairs, 
six with {\it UU} pairs, and eight with {\it VV} pairs.
The sample also includes four sources whose X-ray absorber may be either Compton-thin or Compton-thick (see Section \ref{subsec_NH}): each of them is associated with either a $VV$ pair or a $VL$ pair.
The types of (\NHns,\NHIns) pairs associated with the sources of the correlation sample are listed in Table \ref{tab_NHI_NH}, Column 6.

In order to deal with the four ambiguous Compton-thin/Compton-thick sources, we considered two limiting cases 
in our correlation study. 
In the first case, we assumed these four sources to be all Compton-thin, and hence we associated them with {\it VV} pairs.
This choice yielded a correlation sample composed of 12 sources with (\NHns,\NHIns) pairs of values ({\it VV} pairs) 
and 10 sources with (\NHns,\NHIns) pairs including upper limits (either {\it VU} or {\it UU} pairs), for a total of 22 sources with 23 (\NHns,\NHIns) pairs\footnote{One source is associated with two {\it VV} pairs; see Section 4.2.} of estimates of either the {\it VV}, the {\it VU}, or {\it UU} type; hereafter, this sample is referred to as the {\it estimate sample} $E'$. 
The data in sample $E'$ have only one type of censoring (i.e., the sample pairs include only values and upper limits; there is no mix of upper and lower limits).

In the second case, we assumed the above four sources to be all Compton-thick, and associated them with 
the corresponding {\it VL} pairs.
This left us with a correlation sample composed of three main subsamples:
a subsample of eight sources with (\NHns,\NHIns) {\it VV} pairs, a subsample of 10 sources including upper limits 
(either {\it VU} or {\it UU} pairs), and a subsample of four sources including lower limits ({\it VL} pairs). 
However, because of difficulties in performing the correlation analysis on samples including data with two types of 
censoring (i.e., both upper and lower limits; see Section \ref{sec_correlation} for details), 
we chose to drop the four {\it VL} pairs from the correlation sample. 
The reason why we dropped this subsample of pairs is that this is the smaller of the two subsamples of censored 
data pairs. 
Dropping the subsample that includes the {\it VU} and {\it UU} pairs would have lowered the total number of sources.
Our choice left us with a sample including eight sources with (\NHns,\NHIns) {\it VV} pairs, and  10 sources with (\NHns,\NHIns) pairs of estimates of either the {\it VU} or the {\it UU} type, for a total of 18 sources with 19 (\NHns,\NHIns) pairs of estimates of either the {\it VV}, the {\it VU}, or the {\it UU} type; hereafter, this sample is referred to as the  
{\it estimate sample} $E''$.
By construction, the data in sample $E''$ have only one type of censoring.

\section{Correlation analysis}
\label{sec_correlation}

\begin{figure*}
\center{
\hbox{
\includegraphics[scale=0.68]{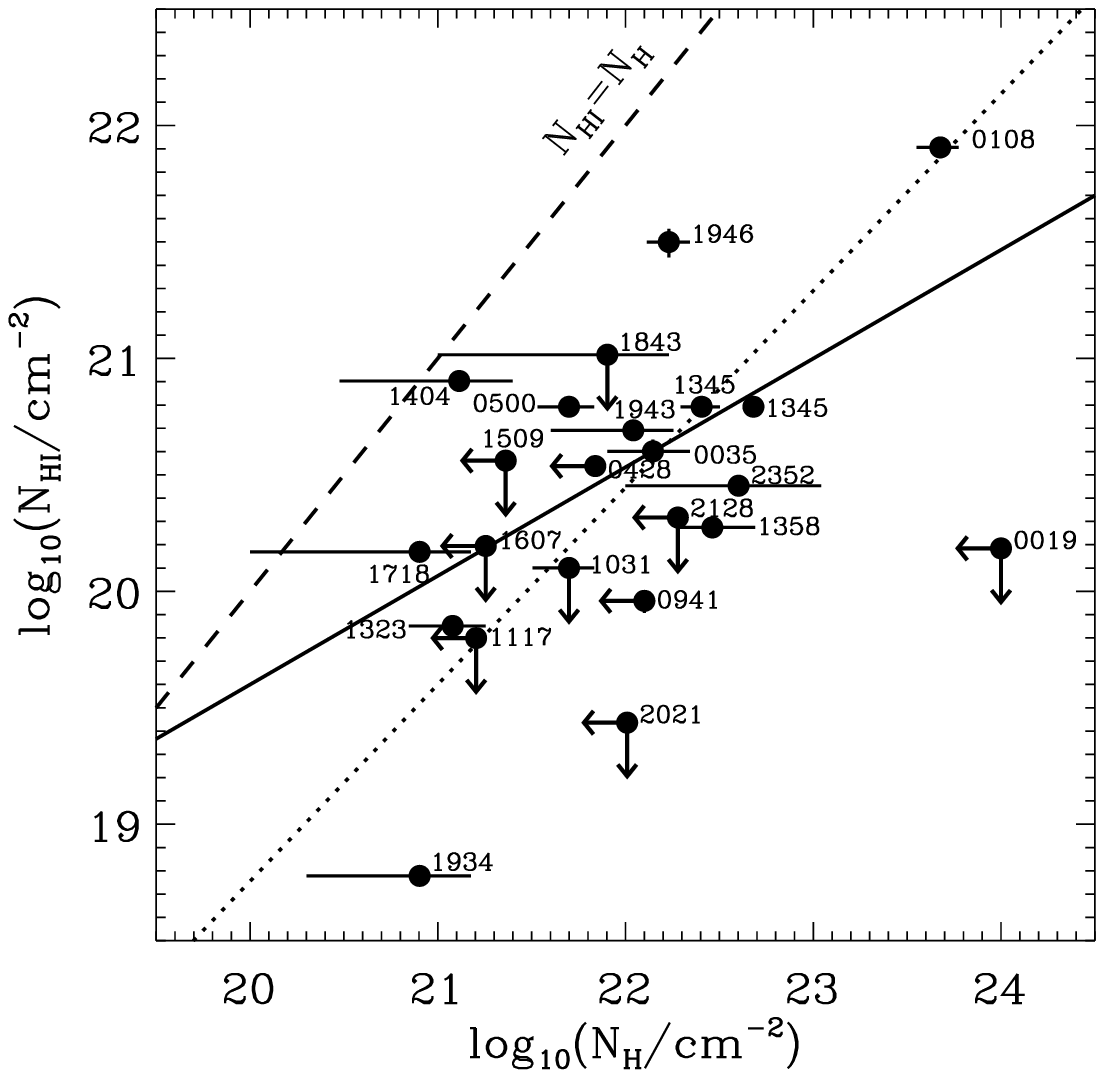}
\includegraphics[scale=0.68]{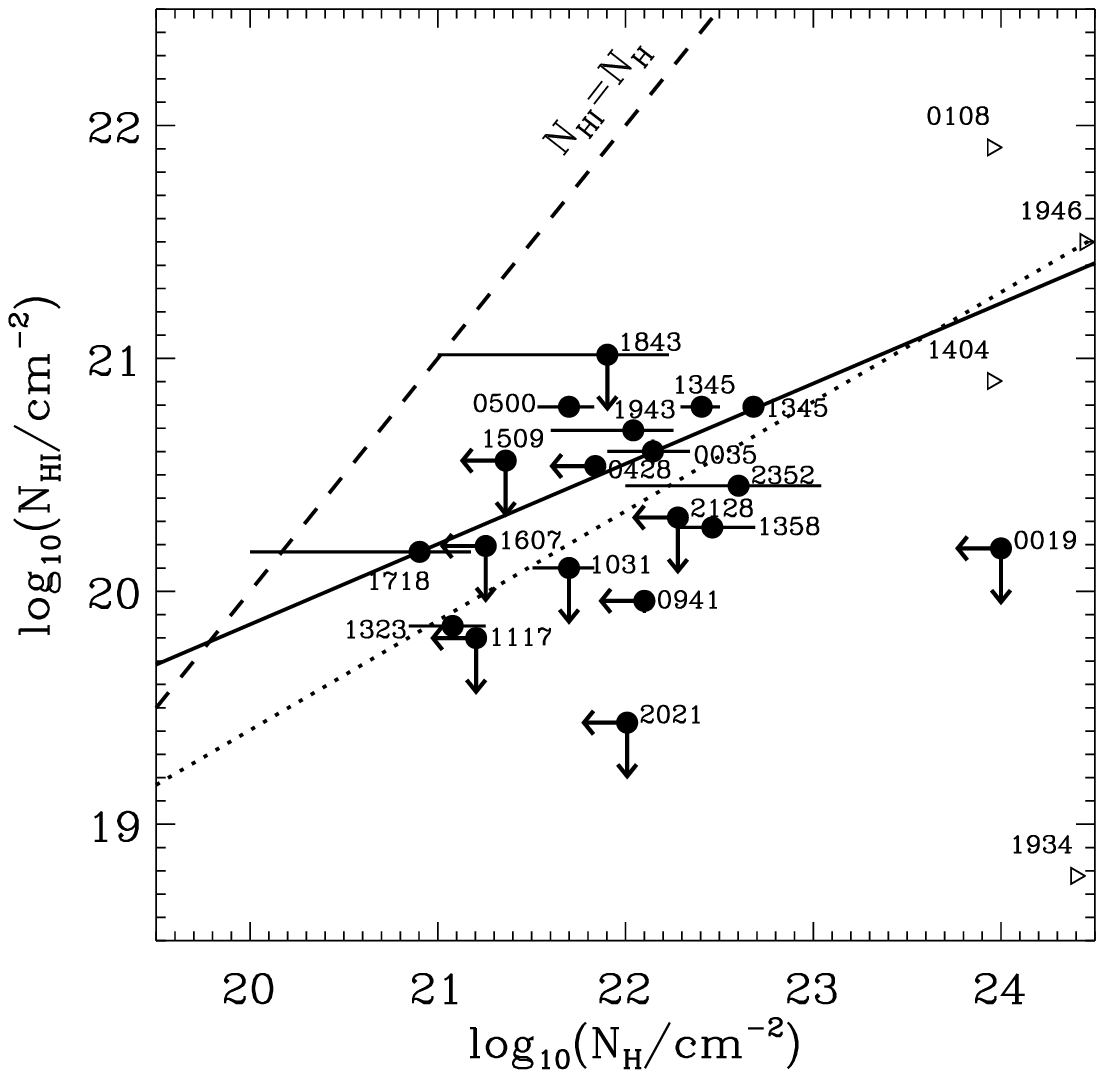}
}
}
\figcaption{Radio column densities (\NHIns) as a function of X-ray column densities (\NHns) for the estimate samples, $E'$ (left) and $E''$ (right).
The solid symbols show the (\NHns,\NHIns) measurements with  1$\sigma$ error bars on \NHns. 
Arrows represent upper limits. 
Lower limits to \NH corresponding to possibly Compton-thick sources are shown in the plot on the right as open triangles, although they were not included in the correlation analysis.
Labels show a shortened version of the source names reported in Table \ref{tab_NHI}.
\NHI was computed by assuming $T_{\mathrm{s}}=100$ K and $C_{\mathrm f}=1$.
Solid line: Akritas-Thiel-Sen regression line, to guide the eye ($\log_{10}$\NHIns$=a+b\log_{10}$\NHns; see Table \ref{tab_corr_est}); dotted line: Schmitt's regression line, for a bin number of 10 (see Section \ref{sec_correlation} for details); dashed line: bisector of the \NHns-\NHI plane. 
\label{fig_corr_est}
}
\end{figure*}

We performed the correlation analysis on both the estimate samples, $E'$ and $E''$, defined in Section \ref{subsec_NH_NHI}.
The results of this analysis are reported in Table \ref{tab_corr_est}, and are discussed below. 
Figure \ref{fig_corr_est} displays the (\NHns,\NHIns) data for these two samples, as well as 
the corresponding linear regression lines, to guide the eye. 

Before discussing these results, we emphasize that correlating \NH with \NHI 
is interesting from the point of view of the physics of the sources, although the estimate of  
\NHI is a combination of the measurement of $\tau_{\rm obs}$ and the assumption on $T_{\rm s}/C_{\rm f}$
(\NHIns$\propto \tau_{\rm obs} \times T_{\rm s}/C_{\rm f}$; see Equation \ref{eq_NHI_approx}).

\begin{deluxetable*}{lccccccccc}
\tabletypesize{\scriptsize}
\tablecaption{Correlation and Regression Analysis for the Estimate Samples, $E'$ and $E''$, by means of 
              Survival Analysis Techniques. \label{tab_corr_est}}
\tablewidth{0pt}
\tablehead{
\colhead{Sample\tnm{a}} & 
\colhead{N$_{\rm data}$} & 
\multicolumn{2}{c}{Generalized Kendall} &
\multicolumn{2}{c}{Generalized Spearman\tnm{d}} &
\multicolumn{2}{c}{Schmitt's Linear Regression\tnm{e}} &
\multicolumn{2}{c}{Akritas-Theil-Sen Linear Regression} \\
\colhead{~~} & 
\colhead{(N$_{\rm sources}$)} &  
\colhead{$z$\tnm{b}} &
\colhead{$P$\tnm{c}} &
\colhead{$\rho$} & 
\colhead{$P$\tnm{c}} &
\colhead{{\it Slope (b)}} & 
\colhead{{\it Intercept (a)}} &
\colhead{{\it Slope (b)}} & 
\colhead{{\it Intercept (a)}}
}
\startdata
$E'$  & 23 (22) & 3.044 & 0.0023 & 0.679 & 0.0014 & 0.820 & 2.42  & 0.467 & 10.3   \\
$E''$ & 19 (18) & 2.615 & 0.0089 & 0.667 & 0.0046 & 0.469 & 10.0  & 0.345 & 13.0  \\
\enddata
\tablenotetext{a}{The samples are defined in Section \ref{subsec_NH_NHI} and in Table \ref{tab_NHI_NH}.}
\tablenotetext{b}{According to the ASURV Rev. 1.3 software manual, $z$ is an estimated function of the correlation and should not be directly compared to the Spearman's correlation $\rho$. The values to be compared are the corresponding probabilities.}  
\tablenotetext{c}{Probability of the null hypothesis of no correlation being true. It is a two-sided significance level: because we are looking a priori for a positive correlation, this value should actually be divided by 2, improving the significance by a factor of 2.}
\tablenotetext{d}{According to the ASURV Rev. 1.3 software manual, the generalized Spearman correlation is not dependable for samples with $N<30$ items, as in our case. In these cases, the generalized Kendall's $\tau$ test should be relied upon. We report it here only for comparison purposes.}
\tablenotetext{e}{For a bin number of 10 for the data set (see Section \ref{sec_correlation} for details).}
\end{deluxetable*}

In the estimate samples, we investigated the correlation by means of survival analysis techniques. 
In particular, we made use of the software package ASURV Rev.\ $1.3$\footnote{\tt http://astrostatistics.psu.edu/statcodes/asurv}
\citep{lavalley1992}, which implements the methods for bivariate problems presented in \citet{isobe1986}.
The generalized Kendall's correlation analysis applied to samples $E'$ and $E''$  
shows that the data are significantly correlated: the probability of no correlation being true is 
$P = 2.3 \times 10^{-3}$ for sample $E'$, and $P = 8.9 \times 10^{-3}$ for sample $E''$.
The generalized Spearman's correlation analysis applied to the same samples confirms the above results: 
$P = 1.4 \times 10^{-3}$ for sample $E'$, and $P = 4.6 \times 10^{-3}$ for sample $E''$.
The reason why the significance of the correlation for sample $E''$ is slightly lower than for sample $E'$ 
is that sample $E''$ does not include, among other sources, the two targets characterized by extreme 
(\NHns,\NHIns) values. 

In order to describe the relationship between \NH and \NHIns, we performed a regression analysis with  
\NHI as the dependent variable. 
The selection criteria of the correlation sample (see Section \ref{sec_sample}) do not introduce 
any bias in this sample; therefore, either \NHI or \NH can formally play the role of the dependent variable. 
Our choice of \NHI as the dependent variable is motivated by the fact that, from a physical perspective, in a given gas distribution, the mass fraction of neutral, atomic gas depends upon the total (i.e., molecular, atomic, and ionized) gas mass through the physical properties of the gas.
As a consequence, for the same gaseous structure, the contribution of \NHI to \NH depends on the physical conditions 
and on the geometrical distribution of the gas (i.e., temperature and covering factor).

According to the generalized Kendall's and Spearman's tests, we could fit a linear relation to the estimate samples $E'$ and $E''$: $\log_{10}$\NHI$=a+b$ $\log_{10}$\NH. 
We first performed a linear fit to the data by means of the ASURV Schmitt's linear regression. 
This method requires a binning of the data set. 
For sample $E'$, by varying the number of data bins from 3 to 15, we obtained slopes in the range  
$b = 0.51-0.99$; for sample $E''$, by varying the number of data bins from 3 to 15, we obtained slopes in the range 
$b = 0.41-0.57$.

For both $E'$ and  $E''$, we show the representative results for the case of 10 bins (dotted lines in Figure \ref{fig_corr_est}).
Because the results of Schmitt's regression analysis are sensitive to the bin size for small data sets,
we performed an additional estimation of the slope with the Akritas-Theil-Sen estimator \citep[e.g.,][]{akritas1995,feigelson2012}, that is  
implemented in the function {\it cenken} in the CRAN package NADA within the R statistical software environment.
With this method, we found regression-line slopes $b=0.47$ and $b=0.35$ for samples $E'$ and $E''$, respectively (solid lines in Figure \ref{fig_corr_est}).
We adopted the slopes derived with the Akritas-Theil-Sen estimator as our trustworthy estimates because of their independence of 
the binning of the data set.

We note that the Akritas-Theil-Sen method applied to samples $E'$ and $E''$ after switching the roles of the variables (i.e. by assuming \NHI as the independent variable) returned slopes $B=0.76$ and $B=1.10$, respectively, for the regression line $\log_{10}$\NH$=A+B$ $\log_{10}$\NHI, in agreement with our previous results 
\citep{ostorero2009,ostorero2010}.

\section{Discussion}
\label{sec_discussion}

With the survival analysis methods described in Section \ref{sec_correlation}, 
we found \NH and \NHI in samples $E'$ and $E''$ to be significantly correlated and to be related to each other through the relationship 
\NHIns$\propto$\NHns$^b$, with $b=0.47$ and $b=0.35$.
Neither the uncertainties on the regression parameters nor the goodness of the 
fit could be evaluated. However, a visual inspection of this data set suggests a dispersion larger than the typical 
uncertainties on \NHI (see Figure \ref{fig_corr_est}).

This fact is supported by the regression analysis of the subsamples that we drew from $E'$ and $E''$ by selecting the detections only.
These two subsamples, hereafter referred to as 
{\it detection samples} $D'$ and $D''$, respectively, display a significant \NH - \NHI correlation according to both Pearson's and Kendall's correlation analysis.
However, the best-fit linear relation turns out not to be a good description of the data ($\chi^2_{\mathrm{red}}\sim 10$):
the dispersion of the data is clearly larger than the typical uncertainties.\footnote{For the detection samples, we performed the regression analysis on a data set where \NH is the dependent variable (i.e., $\log_{10}$\NHns$=A+B$ $\log_{10}$\NHIns), because the uncertainties are available for all the \NH measurements, whereas they are available for a minority of \NHI measurements only. 
Because the uncertainties on \NHI are typically smaller than those on \NH, the low significance of the linear fit also holds for the reverse relation.}

This evidence suggests that the observed \NH - \NHI relation is the two-dimensional projection of a 
multi-dimensional relation, where $T_{\rm s}/C_{\rm f}$ is a variable rather than a parameter. 
In fact, as we show below, $T_{\rm s}/C_{\rm f}$ is the most relevant additional variable in the \NH - \NHI
relation. There are only two other possible variables; however, they either have a modest impact or, ultimately, depend 
on $T_{\rm s}$: 
(i) the abundance of chemical elements that enters the photoionization 
cross-section of the X-ray absorbing gas, ultimately affecting the \NH estimates in the X-ray spectral fitting; and 
(ii) the amount of ionized and molecular hydrogen, \HII and H$_2$.

As for item (i), typical cross-sections adopted in the spectral analysis are based on either solar or ISM abundances. 
These different assumptions lead to variations of the cross-section by a factor of a few \citep[e.g.,][]{ride1977,wilms2000}, implying corresponding variations of the \NH estimates by a factor of a few.
This variation is comparable to the average \NH uncertainty. 

As for item (ii), the abundance of \HII and H$_2$ is an unknown that, in principle, can contribute both to the 
\NH - \NHI offset and to the spread. This abundance is ultimately set by the kinetic temperature of the gas, 
$T_{\rm k}$.
If all the sources were characterized by a similar kinetic temperature of the absorber, the fractions of 
\HII and/or H$_2$  would be comparable in different sources and would only contribute to the offset.
On the other hand, if the kinetic temperature were significantly different in different sources, 
the abundance of \HII and H$_2$ would also 
contribute to the spread, and \NH and \NHI might even be uncorrelated in the case of extreme temperature fluctuations.
However, we do find a correlation between \NH and \NHIns: we can thus exclude extreme fluctuations 
in the kinetic temperature, $T_{\rm k}$.

Similarly, the ratio $T_{\rm s}/C_{\rm f}$ might also significantly fluctuate from source to source; 
for example, it is seen to vary by a factor of at least $\sim 170$ in damped Ly-$\alpha$ systems, 
where \HI column densities are known \citep{curran2013}.
Clearly, extreme fluctuations of $T_{\rm s}/C_{\rm f}$ from source to source might also erase 
the \NH - \NHI correlation.
As in the case of $T_{\rm k}$, the correlation we found excludes extreme fluctuations in $T_{\rm s}/C_{\rm f}$.
Because $T_{\rm k}$ and $T_{\rm s}$ are related to each other, we can ultimately ascribe the correlation spread to fluctuations of $T_{\rm s}/C_{\rm f}$ about the assumed value.

To sum up, for a given source, the \NHI estimate from a spatially unresolved measurement of $\tau_{\rm obs}$
must assume a value for the ratio $T_{\rm s}/C_{\rm f}$; typically, $T_{\rm s}/C_{\rm f}$=100 K is assumed. A different assumption about $T_{\rm s}/C_{\rm f}$  clearly leads to a different \NH - \NHI offset.
When we look for a correlation between X-ray and radio absorption in a sample of sources, we must also assume a $T_{\rm s}/C_{\rm f}$ ratio for each source. The simplest assumption is to assign the same ratio to all the sources. A posteriori, this 
assumption appears to be reasonable because we do find a correlation. 
However, this correlation shows a non-null spread, suggesting that each individual source might actually have a $T_{\rm s}/C_{\rm f}$  ratio slightly different from the value assumed  for the entire sample. By estimating the spread, one can, in principle, estimate the fluctuations of the $T_{\rm s}/C_{\rm f}$  ratio of the individual sources about the value assumed for the entire sample.
As a proof of concept, we estimate the spread of the detection samples $D'$ and $D''$ in Section \ref{sec_bayes}.

\subsection{Quantifying the spread of the \NH - \NHI correlation}
\label{sec_bayes}

In order to simultaneously derive the correlation parameters of the \NH - \NHI relation and 
quantify, for a given \NHns, the intrinsic scatter of \NHI that might be due to the fluctuation of 
the ratio $T_{\rm s}/C_{\rm f}$ of the individual sources about a mean value,
it is appropriate to resort to a Bayesian analysis.
As a proof of concept, we performed a Bayesian analysis of the detection samples $D'$ and $D''$.
We made use of the code APEMoST\footnote{Automated Parameter Estimation and Model Selection Toolkit; 
{\tt http://apemost.sourceforge.net/}, 2011 February.}, which was developed by J.~Buchner and M.~Gruberbauer \citep{gruberbauer2009}
and is suitable for non-censored data sets.
A more sophisticated code, able to perform the Bayesian analysis on samples that include double-censored data (as our samples 
$E'$ and $E''$), could be constructed based on the model developed by \citet{kelly2007}; however, this implementation is beyond 
the scope of the present paper. 

Although samples $D'$ and $D''$ are biased, because they do not include non-detections, 
the results presented below are useful to illustrate how the intrinsic spread of the \NH - \NHI relation 
can be quantified and interpreted. An additional advantage of the Bayesian analysis over the frequentist analysis 
is the possibility to take the uncertainties on both variables into account, even when the uncertainties are  
asymmetric.

Our data set is $DS$=\{$\log_{10}$\NHns${^k}$, $\log_{10}$\NHIns${^k}$, $\mathbf{S}^k$\}, where   
$\mathbf{S}^k=\{\sigma^k_{+}, \sigma^k_{-}, \epsilon^k_{+}, \epsilon^k_{-}\}$ is the vector of 
the upper and lower uncertainties on the $k$-th measures \NHns$^k$ and \NHIns$^k$.
The uncertainties on the \NHI values, for a given $T_{\rm s}/C_{\rm f}$, 
are available for two measurements only: the relative uncertainties are equal to 12\% and 14\%, respectively.
Assuming that the remaining \NHI measurements are affected by comparable uncertainties,
for the purpose of the Bayesian analysis only, to each of them we associated a conservative, 
relative uncertainty of 15\%.
Because the uncertainties on \NHI are much smaller than those on \NHns, all the \NHI uncertainties have 
negligible effects on our results.

Given our data set $DS$, we determined the multi-dimensional probability density function (PDF) of the parameters 
$\theta=\{a, b, \sigma_{\rm int,NHI}\}$, where $a$ and $b$ are the parameters of the correlation (i.e., our model $M$): 
\begin{equation}
\log_{10}N_{\rm HI} = a + b \log_{10}N_{\rm H} \pm \sigma_{\rm int,NHI}, 
\end{equation}
and $\sigma_{\rm int,NHI}$ is the intrinsic spread of the dependent variable.

\begin{figure*}
\center{
\hbox{
\includegraphics[scale=0.4]{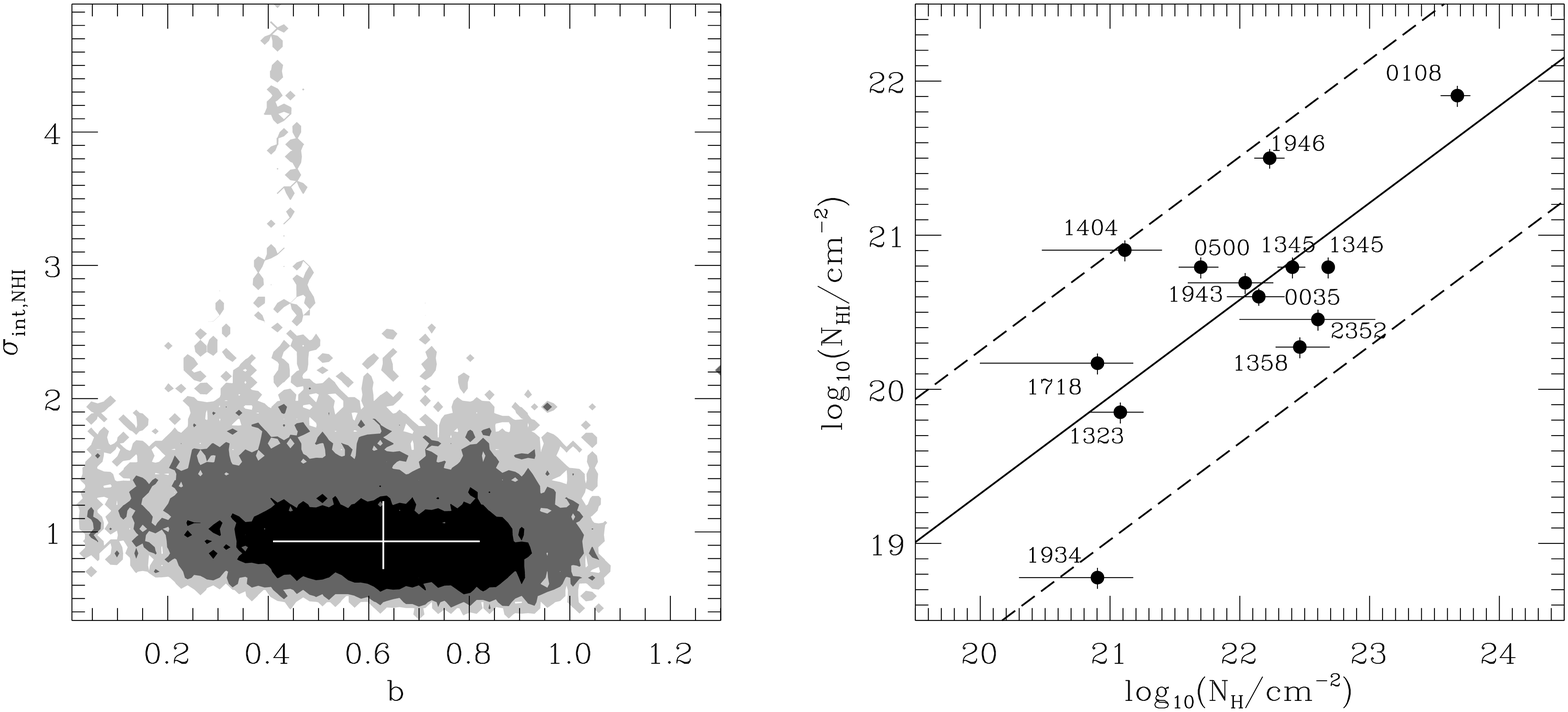} 
}
\hbox{
\includegraphics[scale=0.4]{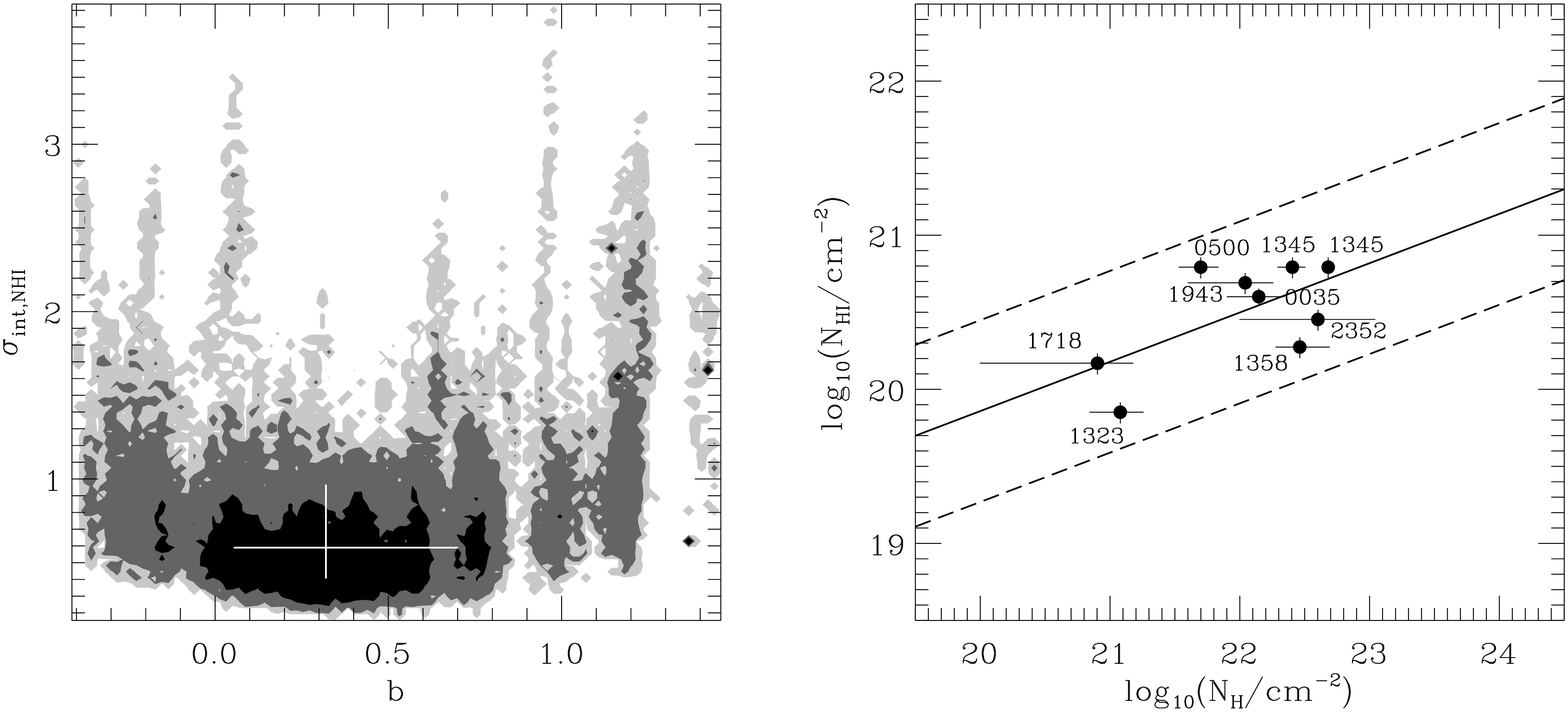} 
}
}
\figcaption{Bayesian analysis applied to sample $D'$ (top) and $D''$ (bottom), with \NH as the independent variable and \NHI as dependent variable. 
The left panels show the marginalized PDFs of the parameters $b$ and $\sigma_{\rm int,NHI}$.
Black, gray, and light-gray shaded regions correspond to the 68.3, 95.4, and 99.7\% confidence levels, respectively.
The crosses show the median values and their marginalized 1$\sigma$ uncertainty. The right panels show the \NH - \NHI correlation: 
the solid symbols show the (\NHns,\NHIns) measurements with their 1$\sigma$ error bars (relative uncertainties of 15\% were assumed 
for the \NHI values whose uncertainty was not available in the literature; see text for details);
the solid, straight line is the \NH - \NHI relation $\log_{10}$\NHIns$=a+b\log_{10}$\NHns; 
the dashed lines show the $\pm\sigma_{\rm int,NHI}$ standard deviation of the relation; the parameters $a$, $b$, and $\sigma_{\rm int,NHI}$ are listed in Table \ref{tab_corr_det_bayes_yx}.
\label{fig_corr_det_bayes_yx}
}
\end{figure*}

In our analysis with APEMoST, we assumed independent flat priors for parameters $a$ and $b$.
For the internal dispersion $\sigma_{\rm int,NHI}$, which is a positive parameter,
we assumed
\begin{equation}
p(\sigma_{\rm int,NHI}\vert M) = {\mu^r\over \Gamma(r)}  x^{r-1} \exp(-\mu x)  \, ,
\end{equation}
where $x=1/\sigma_{\rm int,NHI}^2$, and $\Gamma(r)$ is the Euler gamma function.
This PDF describes a variate with mean $ r/\mu$, and variance $ r/\mu^2$.
We set $r=\mu=10^{-5}$ to assure an almost flat prior.

We used $2\times10^6$ MCMC iterations to guarantee a fairly complete sampling of the parameter space.
The boundaries of the parameter space were set to $[-1000, 1000]$ for the $a$ and $b$ parameters, and to 
$[0.01, 1000]$ for the $\sigma_{\rm int,NHI}$ parameter. 
The initial seed of the random number generator was set with the {\tt bash} command {\tt \verb"GSL_RANDOM_SEED=$RANDOM"}.

The results of this analysis are reported in Table \ref{tab_corr_det_bayes_yx}, and displayed in 
Figure \ref{fig_corr_det_bayes_yx}.


\begin{deluxetable*}{lcccc}
\tabletypesize{\scriptsize}
\tablecaption{Bayesian Analysis of the Detection Samples, $D'$ and $D''$: 
          Median Fit Parameters of the \NH - \NHI Relation. \label{tab_corr_det_bayes_yx}}
\tablewidth{0pt}
\tablehead{
\colhead{Sample} & 
\colhead{N$_{\rm data}$ (N$_{\rm sources}$)} & 
\colhead{$a$} & 
\colhead{$b$} &
\colhead{$\sigma_{\rm int,NHI}$}
}
\startdata
$D'$ & 13 (12) & $6.7^{+4.8}_{-4.2}$ & $0.63^{+0.19}_{-0.22}$ & $0.92^{+0.30}_{-0.21}$ \\
~~\\
$D''$ & 9 (8) & $13.5^{+5.8}_{-8.4}$  & $0.32^{+0.38}_{-0.27}$ & $0.59^{+0.38}_{-0.18} $\\ 
\enddata
\tablenotetext{}{{\bf Note. } The uncertainties are the marginalized 68.3\% confidence intervals.}
\end{deluxetable*}

Our analysis applied to sample $D'$ shows that, for any given value of $\log_{10}$\NHns, $\log_{10}$\NHI takes the 
value $\log_{10}$\NHI$=a+b\log_{10}$\NH$\pm\sigma_{\rm int,NHI}$ 
(with $a=6.7^{+4.8}_{-4.2}$, $b=0.63^{+0.19}_{-0.22}$, and 
$\sigma_{\rm int,NHI}=0.92^{+0.30}_{-0.21}$), with a 68\% probability; 
this value of the intrinsic spread implies that, for any given \NHns, 
the corresponding \NHI falls within a factor of $\simeq 8$ from the mean relation at the 68\% confidence level.

As for the smaller sample $D''$, our analysis yielded $b=0.32^{+0.38}_{-0.27}$ and a smaller intrinsic spread, 
$\sigma_{\rm int,NHI}=0.59^{+0.38}_{-0.18}$, which implies that, for any given \NHns,
the corresponding \NHI falls within a factor of $\simeq 4$ from the mean relation with 68\% probability.
Because $D''$ was obtained from $D'$ by removing the four ambiguous Compton-thin/thick sources, 
this means that those sources were responsible for the larger spread on \NHI that we found 
for sample $D'$ (compare the top and bottom panels on the right-hand-side of Figure \ref{fig_corr_det_bayes_yx}).

Overall, for the detection samples $D'$ and $D''$, the value of the regression line's slope is $b\simeq 0.3-0.6$,
consistent with the results of the regression analyses presented in Section \ref{sec_correlation} for the estimate samples $E'$ and $E''$.  

We discuss the possible implications of our exercise on the evaluation of the spread that might be due to $T_{\rm s}/C_{\rm f}$ fluctuations in Section \ref{sec_implications}.

\subsection{Implications of the \NH - \NHI correlation and its spread}  
\label{sec_implications} 

The \NH - \NHI correlation that we found in Section \ref{sec_correlation} suggests a physical connection between the X-ray and \HI absorbers: the gas responsible for the X-ray obscuration and the \HI absorption in compact radio galaxies may be part of the same, possibly unsettled, gaseous structure that extends over scales of a few hundred parsecs.

This scenario, also supported by recent X-ray and \HI observations of a composite sample of compact radio sources \citep{glowacki2017},
is corroborated by the X-ray absorption properties of the full X-ray emitting GPS/CSO sample:
the mean total hydrogen column density of this sample (see Table \ref{tab_NH}) varies from 
\NH $\simeq (0.8-1) \times 10^{22}$ cm$^{-2}$ ($\sigma_{N_H}\simeq 0.2-0.3$ dex) to 
\NH $\simeq 3 \times 10^{22}$ cm$^{-2}$ ($\sigma_{N_H}\simeq 1$ dex), depending on whether absorbers that are not unambiguously Compton-thin are considered to be Compton-thin or Compton-thick sources, respectively.
These values are consistent with a picture where the absorption of the X-rays from compact radio galaxies of the GPS/CSO type is 
comparable to the X-ray absorption in the extended FR-I and the unobscured FR-II radio galaxies.
Together with the \NH - \NHI correlation, which points toward a physical connection between the X-ray and radio absorbers, 
this evidence suggests that, in GPS/CSOs, the X-ray absorbing gas is located on scales larger than those of the parsec-scale, dusty tori typically invoked in AGN unification schemes.
Such a scenario would imply that either ``standard'' dusty tori are not present in compact radio galaxies, 
or that the dominant contribution to the X-ray emission of GPS/CSOs does not originate in the accretion disc, 
but rather in the larger-scale jet/lobe components.

Even though the \NH - \NHI correlation is statistically significant, the correlated data set is affected by a large intrinsic spread. For the detection samples, we could quantify the spread by means of a Bayesian analysis 
(Section \ref{sec_bayes}).
This spread could potentially provide us with interesting constraints on the properties of the neutral
hydrogen in the ISM of the host galaxies of compact radio sources.

The  \NH - \NHI correlation that we found actually reflects a correlation between \NH and $\tau_{\rm obs}$.
Indeed, \NHIns$\propto T_{\rm s}/C_{\rm f} \times \tau_{\rm obs}$, with $\tau_{\rm obs} \equiv \int{\tau_{\rm obs}(v)\,{\mathrm d}v}$ (see Equation
(\ref{eq_NHI_approx})); in our case, the only observable is the 
velocity-integrated optical depth of the absorption line, $\tau_{\rm obs}$, 
because we only considered spatially unresolved observations ($C_{\rm f}=1$) and 
we assumed $T_{\rm s}=100$ K for the absorbing gas, so we obtained a ratio of $T_{\rm s}/C_{\rm f}=100$~K 
for all the sources.

However, the ratio $T_{\rm s}/C_{\rm f}$ is seen to vary by a factor of at least $\sim 170$ 
(from 60 K to 9950 K) in damped Ly-$\alpha$ systems, where \NHI column densities are known \citep{curran2013}; 
furthermore, the analysis of the compact quasar PKS 1549$-$79 suggests that $T_{\rm s}>3000$ K in this 
source \citep{holt2006}; therefore, the assumption of a common value $T_{\rm s}/C_{\rm f}=100$ K for the sources 
analized here is not well-justified.

If \NH and \NHI were intrinsically tightly correlated, the assumption of a constant $T_{\rm s}/C_{\rm f}$ would be responsible for the \NHI spread that we measured.
As an example, when the intrinsic spread of \NHI is $\sigma_{\rm int,NHI}=0.92$ (as for sample $D'$, see 
Table \ref{tab_corr_det_bayes_yx}), for any given \NHns, the corresponding observed \NHI falls within a factor 
of $\simeq 8$ from the mean relation.
However, if the correct value of \NHI actually lies on the mean relation, its observed deviation derives from the incorrect 
assumption $T_{\rm s}/C_{\rm f}=100$ K and the factor of $\simeq 8$ has to be associated to the fluctuations of $T_{\rm s}/C_{\rm f}$ about 100 K; in other words, this ratio is expected to be in the range $T_{\rm s}/C_{\rm f}=12-832$ K, at the 68\% confidence level.

\begin{figure}[!b]
\includegraphics[scale=0.36]{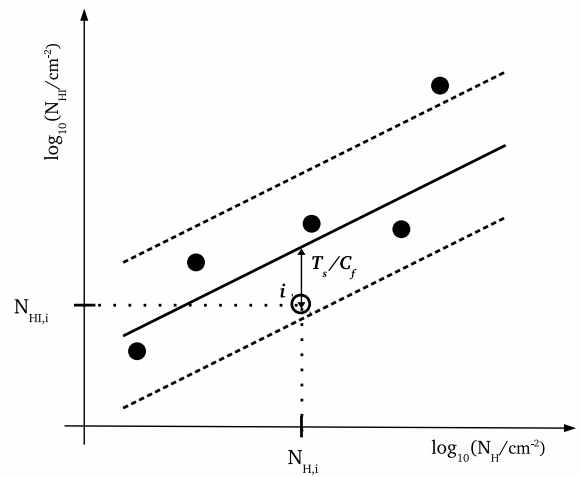}   
\caption{If a true \NH - \NHI correlation exists (solid line), the observation of a source $i$ off the correlation is due to a deviation of $T_{\rm s}/C_{\rm f}$ from the value assumed for the entire \NHI sample. Specifically, a source  appears below the mean relation when the value of $T_{\rm s}/C_{\rm f}$ assumed for the sample is too low: the value  of $T_{\rm s}/C_{\rm f}$ that is appropriate for source $i$ and brings it back on the mean relation is proportionally larger, as indicated by the arrow.}
\label{fig_spread}
\end{figure}

Therefore, we can, in principle, use our Bayesian result to forecast the true \NHI value and constrain the 
$T_{\rm s}/C_{\rm f}$ fluctuation: 
for any given \NHns, the corresponding true \NHI value lies on the mean relation; by comparing the true value with the observed value, we can infer the deviation of the $T_{\rm s}/C_{\rm f}$ value from the assumed 100 K. 
This argument is sketched in Figure \ref{fig_spread}.

In our sample, the \HI observations do not spatially resolve the source and the \HI is observed only in absorption.
Therefore, $T_{\rm s}$ and $C_{\rm f}$ cannot be disentangled, and the estimate of $T_{\rm s}/C_{\rm f}$ does not enable to constrain the spin temperature of the gas.
On the other hand, if \HI measures of a sufficiently large sample of sources 
were based on high angular resolution observations, constraining $T_{\rm s}$ would, in principle, be possible.
Indeed, this kind of observation enables to locate the \HI absorber; the absorption profiles are derived only for the 
source region covered by the absorber, and the condition $C_{\rm f}=1$ is readily justified.
Moreover, Equations (\ref{eq_NHI}) and (\ref{eq_NHI_approx}) are more suitable for resolved observations;
those equations hold under the assumption that the source is homogeneous, as appropriate
when the source fraction considered for the evaluation of \NHI is small.

In conclusion, if a sample of resolved sources confirmed a correlation between \NH and \NHI, the deviation 
between the data-point of an individual source and the mean relation would return an estimate of the  
deviation of the spin temperature $T_{\rm s}$ of the \HI in that source from the $T_{\rm s}$ assumed for the \NHI estimate of the entire sample.

\section{Conclusions}
\label{sec_conclusions}

We performed spatially unresolved \HI absorption observations of a sample of X-ray emitting GPS/CSO galaxies
with the WSRT, in order to improve the statistics of the \NH - \NHI correlation sample and thoroughly investigate 
the possible connection between X-ray and radio absorbers in compact radio galaxies.

We confirmed the significant positive correlation between \NH and \NHI found by \citet{ostorero2010}, 
which implies that GPS/CSOs with increasingly large X-ray absorption have an increasingly larger 
probability of being detected in \HI absorption observations.
More interestingly, this correlation suggests that the gas responsible for the X-ray and radio absorption may be part of the same, possibly unsettled, hundred-parsec scale structure.

For the full, censored data set, the Akritas-Theil-Sen regression analysis yields \NHIns$\propto$\NHns$^b$, 
with $b=0.47$ and $b=0.35$, depending on the subsample.
The correlation displays a large intrinsic spread, which may indicate that the \NH - \NHI relation is not a one-to-one relation; additional variables are involved in the correlation and generate the intrinsic spread of the data set.

The estimate of \NHI relies on the assumed value of the ratio $T_{\rm s}/C_{\rm f}$ between the spin temperature of the absorbing gas and its covering factor. We suggested that the additional variable mostly responsible for the \NHI spread
around the mean \NH - \NHI correlation is the source-to-source fluctuation of $T_{\rm s}/C_{\rm f}$ with respect to the value  assumed for the entire sample.
An estimate of the \NHI spread at fixed \NHns, i.e. $\sigma_{\rm int,NHI}$, can be obtained through a Bayesian analysis.  
As a proof of concept, we performed this analysis on two uncensored subsamples only.
For the larger subsample, we found  $\sigma_{\rm int,NHI}=0.92$, implying that, for any given \NHns, $T_{\rm s}/C_{\rm f}$ is expected to fall within a factor of $\simeq 8$, at the 68\% confidence level, from the value $T_{\rm s}/C_{\rm f}=$ 100 K assumed for the whole sample.
If the existence of an \NH - \NHI correlation were confirmed by high angular resolution observations of large enough samples, we could, in principle, disentangle $T_{\rm s}$ from $C_{\rm f}$ and constrain the deviation of the gas spin temperature, $T_{\rm s}$, from the assumed value. This result would provide an unprecedented piece of information on the physical properties of the \HI absorbing gas.

~~\\
\acknowledgments
L.O. gratefully acknowledges support from the {\it Helena Kluyver} program run by ASTRON/JIVE.
She is grateful to Ravi Sheth and to the Department of Physics and Astronomy of the University of Pennsylvania for support during the early stage of the project. 
She acknowledges funding  from the University of Torino and Compagnia di San Paolo through 
the {\it Strategic Research Grant: Origin and Detection of Galactic and Extragalactic Cosmic Rays}.
L.O. and A.D. acknowledge the INFN grant INDARK and the grant PRIN 2012 {\it Fisica Astroparticellare
Teorica} of the Italian Ministry of University and Research.
L.O., A.D., and A.L. are grateful to the Harvard-Smithsonian Center for Astrophysics for their kind hospitality. 
A.S. was supported by NASA through contract NAS8-03060 to the Chandra X-ray Center and NASA grants
GO1-12145X, GO4-15099X. {\L}.S. was supported by Polish NSC grant DEC-2012/04/A/ST9/00083.
R.M. gratefully acknowledges support from the European Research Council under the European Union's Seventh Framework Programme 
(FP/2007-2013) /ERC Advanced Grant RADIOLIFE-320745.
We are grateful to Gyula J\'ozsa for his support during the observing runs with the WSRT and for 
providing us with the data-cubes of our target sources.
We thank Johannes Buchner and Michael Gruberbauer for developing their superb
code APEMoST and making it available to the community (apemost.sourceforge.net).
Stefano Andreon is acknowledged for a very stimulating seminar on Bayesian statistics applied to astrophysics.
We thank Guido Risaliti for stimulating discussions on AGN obscuration.
We gratefully acknowledge an anonymous referee for a very careful revision of 
the paper and for constructive suggestions.
The Westerbork Synthesis Radio Telescope is operated by the ASTRON (Netherlands Institute for Radio 
Astronomy) with support from the Netherlands Foundation for Scientific Research (NWO).
This research has made use of the NASA/IPAC Extragalactic Database (NED), which is operated by the
JPL, CalTech, under contract with NASA.
~~\\
~~\\
{\it Facility:} WSRT\\
{\it Software:} MIRIAD \citep{sault1995}, ASURV \citep{lavalley1992}, APEMoST \citep{gruberbauer2009}

\appendix
All the \NHI and \NH estimates currently available for the 27 GPS/CSOs of the sample described in Section \ref{sec_sample} 
are reported in Tables \ref{tab_NHI} and \ref{tab_NH}, respectively, together with the source properties that are most relevant to the present study. 
In these tables, the 22 sources of the  correlation sample  and the corresponding \NHI and \NH estimates that we 
used for the correlation analysis are highlighted with boldface fonts.  
The criteria that we adopted to select the above column density estimates among all the available estimates 
are described in Appendices \ref{subsec_NHIapp} and \ref{subsec_NHapp}.

~~\\

\section{\NHI estimates}
\label{subsec_NHIapp}
Table \ref{tab_NHI} lists the names of the 27 GPS/CSOs of the sample described in Section \ref{sec_sample} (Column 1), their redshift (Column 2), their radio spectral and morphological classification as GPS and/or CSO (Column 3), their \NHI estimate (Column 4), the type (low/high-angular resolution) of \HI absorption observation from which the \NHI estimate was derived, with possible estimates of the covering factor of the absorbing gas (Column 5), the label associated with the absorption line(s) the \NHI estimate refers to (Column 6), the width of the absorption line (Column 7) and its peak optical depth (Column 8), the instrument used for the \HI observation (Column 9), and the reference for the \HI measurement (Column 10).

In this table, the 22 sources of the  correlation sample  and the corresponding \NHI estimates that we 
used for the correlation analysis are highlighted with boldface fonts.  
The criteria that we adopted to select the column density estimates highlighted in Column 4 
among all the available estimates are described below.

For three sources of the correlation sample, both low angular resolution \HI absorption measurements 
(unable to spatially resolve the source) and high angular resolution \HI absorption measurements (able to spatially 
resolve the source) were available; in Table \ref{tab_NHI}, Column 5, we denoted them by ``U'' (unresolved) and ``R'' 
(resolved), respectively.
For self-consistency of the correlation analysis, we discarded the ``R'' measurements and made use of the 
``U'' measurements only.
For sources that were not detected, 3$\sigma$ upper limits to \NHI were chosen for the analysis.

After applying these criteria, our sample of 22 sources turned out to be composed of two subsamples, in terms of number of available \NHI estimates.
For 10 out of 22 sources, only one unresolved \NHI estimate was available: we took this value  
for our correlation data set.
Conversely, more than one unresolved \NHI estimate was available for any of the remaining 12 sources:
for the purpose of our correlation analysis, we proceeded as follows.

For 5 out of 12 sources, two spectral features (marked with ``I'' and ``II'', respectively, in Table \ref{tab_NHI}, Column 6) 
were detected in the \HI absorption spectra.
In three of them, \NHI was available for each absorption line, but not for the full spectrum: 
we associated with each of these sources a value of \NHI estimated as the sum of the \NHI values 
of each spectral line (this is the value reported in Table \ref{tab_NHI_NH}).
In the other two sources, besides the \NHI of each line, also the total \NHI 
(marked with ``tot.'' in Table \ref{tab_NHI}, Column 6) was available: we associated the latter \NHI value with these sources.

For 7 out of 12 sources, either one or no absorption features were detected in different observations.
When multiple \NHI {\it values} were available for a given source, we chose the most 
recent result. When an \NHI {\it value} and one or more {\it upper limits} to \NHI were available, 
we chose the \NHI value. When different {\it 3$\sigma$ upper limits} to \NHI were available, 
we chose the most stringent one.

\LongTables
\begin{deluxetable*}{lccccccccc}
\tabletypesize{\scriptsize}
\tablecaption{\HI Column Densities and Spectral Parameters.\label{tab_NHI}}
\tablewidth{0pt}
\tablehead{
\colhead{Source Name} &
\colhead{$z_{\mathrm {opt}}$} &
\colhead{GPS/CSO\tnm{a}} &
\colhead{\NHI} &
\colhead{Res./Unres.} &
\colhead{Line nr.} &
\colhead{$\Delta V$} &
\colhead{$\tau_{\mathrm {peak}}$} &
\colhead{Instrument} &
\colhead{References} \\
\colhead{B1950} &
\colhead{~~} &
\colhead{~~} &
\colhead{\big($10^{20}\frac{T_{\mathrm s}}{100 \mathrm{\,K}}$ \percmq \big)} &
\colhead{(and $C_{\mathrm f}$)} &
\colhead{~~} &
\colhead{(km \persec)} &
\colhead{($10^{-2}$)} &
\colhead{~~} &
\colhead{~~} \\
\colhead{(1)} &
\colhead{(2)} &
\colhead{(3)} &
\colhead{(4)} &
\colhead{(5)} &
\colhead{(6)} &
\colhead{(7)} &
\colhead{(8)} &
\colhead{(9)} &
\colhead{(10)} \\
}
\startdata
{\bf 0019$-$000} & 
0.305 		 & 
GPS   		 & 
\nodata 	& 
U       	& 
\nodata 	& 
\nodata 	& 
\nodata 	& 
WSRT    	& 
P03   		\\                  
~~~     	& 
~~~     	& 
~~~     	& 
$\mathbf{<1.5}$\tnm{b,d} &
U       	& 
\nodata 	& 
\nodata 	& 
$<0.84$\tnm{b} 	& 
WSRT   	 	& 
($\star$) 	\\           
0026$+$346 	& 
0.517   	& 
GPS,CSO 	& 
\nodata 	& 
\nodata 	& 
\nodata 	& 
\nodata 	& 
\nodata 	& 
WSRT    	&  
($\star$) 	\\ 
{\bf 0035$+$227} & 
$0.096\pm0.002$  & 
CSO      	& 
$\mathbf{3.99\pm0.49}$ & 
U		& 
\nodata     	& 
$114\pm10$     & 
$1.81\pm0.14$   &
 WSRT       	& 
($\star$)   	\\
{\bf 0108$+$388} & 
0.66847 	& 
GPS,CSO  	& 
80.7    	& 
U       	& 
\nodata  	& 
$94\pm10$ 	& 
$44\pm4$    	& 
WSRT       	& 
C98 		\\                   	
~~~    		& 
~~~    		& 
~~~      	&
$\mathbf{80.5}$ & 
U            	& 
\nodata  	& 
100         	& 
43.7            & 
WSRT          	& 
O06 		\\                   	
0116$+$319 	& 
0.06    	& 
GPS,CSO   	& 
$10.8$          & 
U            	& 
II       	& 
$153\pm6$      & 
$3.7\pm0.1$     & 
VLA   		& 
v89 \\                   	
~~~        	&
~~~     	& 
~~~   	 	& 
$12.2\pm0.14$   & 
U            	& 
tot.     	&   
7.6, 153       & 
3.0, 3.8        & 
Arecibo         & 
G06 		\\    
{\bf 0428$+$205} & 
0.219  		& 
GPS,CSO    	&  
$\mathbf{2.52}$ & 
U               & 
I       	& 
297            & 
$0.46$          & 
WSRT          	& 
V03 \\                 
~~~       	& 
~~~    		& 
~~~        	&  
$\mathbf{0.93}$ & 
U             	& 
II      	& 
247            & 
$0.21$          & 
WSRT          	& 
V03 		\\                      
{\bf 0500$+$019} & 
0.58457 	 & 
GPS,CSO   	 & 
$\mathbf{6.2}$   & 
U            	 & 
tot.     	 & 
$\sim 140$      & 
4                & 
WSRT             & 
C98  \\                    
~~~        	& 
~~~     	& 
~~~       	& 
2.5      	& 
U            	& 
I        	& 
$45\pm9$       & 
$2.7\pm0.3$     & 
WSRT          	& 
C98  \\                     
~~~        	& 
~~~     	& 
~~~       	& 
4.5             & 
U            	& 
II       	& 
$62\pm7$       & 
$3.6\pm0.3$     & 
WSRT            & 
C98  \\                    
0710$+$439 	& 
0.518   	& 
GPS,CSO   	&  
\nodata    	& 
U            	& 
\nodata  	& 
\nodata        & 
\nodata         & 
WSRT         	& 
P03 		\\                
~~~        	& 
~~~     	& 
~~~       	&  
\nodata         & 
U            	& 
\nodata  	& 
\nodata        & 
\nodata         & 
WSRT         	& 
($\star$) 	\\
{\bf 0941$-$080}  & 
0.2281$\pm$0.0013 & 
GPS,CSO   	  & 
$<0.80$\tnm{c}    & 
U            	  & 
\nodata   	  &
\nodata          & 
$<0.44$           & 
WSRT              & 
V03 		 \\                   
~~~        	  & 
~~~     	  & 
~~~       	  & 
$<1.26$\tnm{b}  &
U             	  & 
\nodata   	  & 
\nodata          & 
\nodata           & 
WSRT              & 
P03 \\                  
~~~        	& 
~~~     	& 
~~~       	& 
$\mathbf{0.91\pm0.10}$  & 
U            	& 
\nodata   	& 
$215\pm18$     & 
$0.22\pm0.02$   & 
WSRT         	& 
($\star$) 	\\
{\bf 1031$+$567} & 
0.459   	 & 
GPS,CSO  	 & 
$<0.87$\tnm{c}   & 
U                &
\nodata          & 
\nodata         & 
$<0.48$          & 
WSRT             & 
V03 		\\                   
~~~        	&
~~~             & 
~~~       	& 
$\mathbf{<1.26}$\tnm{b} & 
U           	& 
\nodata   	& 
\nodata        & 
\nodata         & 
WSRT        	& 
P03 		\\    
~~~        	& 
~~~     	& 
~~~       	& 
$<1.8$\tnm{b,d}   & 
U           	& 
\nodata   	& 
\nodata        & 
$<0.99$\tnm{b}  & 
WSRT            & 
($\star$)      \\  
{\bf 1117$+$146} & 
0.362  		& 
GPS,CSO 	&  
$<0.38$\tnm{c}  & 
U           	& 
\nodata    	& 
\nodata        & 
$<0.21$         & 
WSRT            & 
V03 		\\                   
~~~        	& 
~~~    		& 
~~~      	&  
$\mathbf{<0.63}$\tnm{b}  & 
U           	& 
\nodata    	& 
\nodata        & 
\nodata         & 
WSRT         	& 
P03 		\\                      
~~~        	& 
~~~    		& 
~~~      	& 
$<1.1$\tnm{b,d}  &
U           	& 
\nodata    	& 
\nodata        & 
$<0.62$\tnm{b}  & 
WSRT         	& 
($\star$)	\\                  
1245$+$676      & 
0.107   	& 
GPS,CSO 	&  
$6.73$ 		& 
U           	& 
tot.    	& 
\nodata        & 
\nodata        & 
GMRT            & 
S07		\\                   
{\bf 1323$+$321} & 
0.370  		& 
GPS,CSO 	& 
$\mathbf{0.71}$ & 
U           	& 
\nodata    	& 
229             & 
0.17            & 
WSRT         	& 
V03	\\
{\bf 1345$+$125} & 
0.12174 	& 
GPS,CSO 	&  
$\mathbf{6.2}$  & 
U           	& 
tot.      	& 
150       	& 
1.38            & 
Arecibo      	& 
M89 \\ 
~~~        	& 
~~~    		& 
~~~       	&  
$\sim 1.7$      & 
U            	& 
II       	& 
$\sim 2000$    & 
$\sim 0.15$     & 
WSRT         	& 
M03a \\
~~~        	& 
~~~    		& 
~~~       	&  
$\sim 2$        & 
U            	& 
I        	& 
\nodata        & 
$\sim 1$        & 
WSRT         	& 
M03b 		\\ 
~~~       	& 
~~~    		& 
~~~       	&  
$\sim 1$              & 
U            	& 
II       	& 
\nodata        & 
$\sim 0.2$      & 
WSRT         	& 
M03b 		\\ 
~~~        	& 
~~~    		& 
~~~       	&  
$\sim 100$      & 
R ($\sim0.2$) 	& 
I      		& 
150            & 
$\gtsim 60$     & 
VLBI         	& 
M04 		\\                    
~~~       	& 
~~~    		& 
~~~       	& 
440$^\dagger$   & 
R            	& 
I        	& 
\nodata        & 
60              & 
VLBI         	& 
M13 		\\                      
~~~        	& 
~~~    		& 
~~~       	&  
46              & 
R            	& 
II       	& 
\nodata        & 
\nodata            & 
VLBI         	& 
M13 		\\                      
{\bf 1358$+$624} & 
0.431  		& 
GPS,CSO  	& 
$\mathbf{1.88}$ & 
U            	& 
\nodata 	& 
170             & 
0.61            & 
WSRT         	& 
V03 		\\                  
{\bf 1404$+$286} & 
0.07658 	& 
GPS,CSO 	&  
1.83            & 
U            	& 
\nodata & 
256             & 
0.39            & 
WSRT         	& 
V03 		\\                           
~~~        	& 
0.0773  	& 
~~~     	&  
$\mathbf{8.0}$  & 
U            	& 
\nodata 	& 
1800            & 
0.5             & 
WSRT         	& 
O06 		\\ 
{\bf 1509$+$054} & 
0.084  		 & 
GPS,CSO 	 &   
$\mathbf{<3.64}$ &
U           	 & 
\nodata 	 & 
\nodata         & 
$<0.02$          & 
WSRT         	 & 
P03   		\\                   
{\bf 1607$+$268} & 
0.473   	& 
GPS,CSO 	&  
\nodata         & 
U             	& 
\nodata    	& 
\nodata        & 
\nodata         & 
WSRT        	& 
P03   		\\                   
~~~        	& 
~~~     	& 
~~~     	&  
$\mathbf{<1.6}$\tnm{b,d} & 
U             	& 
\nodata    	& 
\nodata        & 
$<0.86$\tnm{b}	 & 
WSRT        	& 
($\star$)	\\                  
{\bf 1718$-$649}  &  
0.0142  	  & 
GPS,CSO 	  &  
$\mathbf{0.703}$  & 
U  		  &  
I  		  & 
43(FWZI)         & 
0.4               & 
ATCA              & 
M14              \\                   
~~~              &  
~~~     	 & 
~~~     	 &  
$\mathbf{0.774}$ & 
U                &  
II               & 
65(FWZI)        & 
0.2              & 
ATCA             & 
M14              \\                   
{\bf 1843+356}  & 
0.764   	& 
GPS,CSO 	& 
$<6.92$\tnm{c}  & 
U 	        & 
\nodata    	& 
\nodata       &
$<3.80$         & 
WSRT        	& 
V03   		\\
~~~        	&
~~~     	& 
~~~     	& 
$\mathbf{<10.4}$\tnm{b,d} & 
U	        & 
\nodata        	& 
\nodata            &
$<5.70$\tnm{b}   & 
WSRT        	& 
V03\tnm{e} 	\\ 
~~~        	& 
~~~     	& 
~~~     	& 
$<11.7$\tnm{b,d}   & 
U	        & 
\nodata        	& 
\nodata           & 
$<6.41$\tnm{b}  & 
WSRT        	& 
($\star$)   	\\
{\bf 1934$-$638} & 
0.18129 	& 
GPS,CSO 	&  
$\mathbf{0.06}$ &
 U              & 
\nodata 	& 
100            & 
0.22            & 
ATCA, LBA   	& 
V00   		\\                  
{\bf 1943$+$546} & 
0.263  		& 
GPS,CSO 	&  
$\mathbf{4.91}$ & 
U              	& 
tot.   		&  
$315$          &  
0.86            & 
WSRT        	& 
V03 		\\
{\bf 1946$+$708} & 
0.101   	& 
GPS,CSO 	&  
$\mathbf{31.6}$ & 
U              	& 
tot.   		&  
357            & 
$5.00\pm0.7$    & 
VLA         	&
P03,P99 \\                   
~~~       	& 
~~~     	& 
~~~     	&  
$[2.5-27.33]$\tnm{f} & 
R ($\sim 1$)   	& 
I-tot. 	& 
$[40.1-357]$   & 
$[3.3-7.0]$   	&
VLBA        	& 
P99   		\\                 
2008$-$068 	& 
0.547   	& 
GPS,CSO 	&  
\nodata 	&
U           	& 
\nodata    	& 
\nodata        & 
\nodata         & 
WSRT        	& 
($\star$) 	\\                  
{\bf 2021$+$614} & 
0.227   	& 
GPS,CSO 	& 
$<0.24$\tnm{c}  & 
U           	& 
\nodata 	& 
\nodata        & 
$<0.13$         & 
WSRT        	& 
V03   		\\
~~~        	& 
~~~     	& 
~~~     	& 
$<0.398$\tnm{b} & 
U           	& 
\nodata 	& 
\nodata        & 
\nodata         & 
WSRT        	& 
P03   		\\
~~~        	& 
~~~     	& 
~~~     	& 
$\mathbf{<0.27}$\tnm{b,d}     & 
U           	& 
~~~     	& 
~~~            & 
$<0.15$\tnm{b}  & 
WSRT        	& 
($\star$) \\
{\bf 2128$+$048} & 
0.99   		& 
GPS,CSO 	& 
$\mathbf{<2.08}$\tnm{b,d} & 
U           	& 
\nodata   	& 
\nodata        & 
$<1.14$\tnm{b}  & 
WSRT        	& 
($\star$) 	\\                            
{\bf 2352$+$495} & 
0.2379 		& 
GPS,CSO 	& 
$\mathbf{0.28}$ & 
U               & 
I    		& 
13              & 
1.16            & 
WSRT            & 
V03 		\\                         
~~~        	& 
~~~    		& 
~~~      	& 
$\mathbf{2.56}$ & 
U             	& 
II   		& 
82             & 
1.72            & 
WSRT         	& 
V03 		\\                      
~~~        	& 
~~~    		& 
~~~      	& 
$0.91\pm0.13$\tnm{f}   & 
R($\sim$0.2)    & 
I    		& 
$13\pm2$       & 
\nodata         & 
VLBA         	& 
A10 		\\                      
~~~        	& 
~~~    		& 
~~~      	& 
$6.5\pm0.3$\tnm{f,$\dagger$} & 
R($\sim$0.2)    & 
II   		& 
$85\pm4$       & 
\nodata         & 
VLBA         	& 
A10 		\\                      
\enddata
\tablenotetext{}{{\bf Note. }The \NHI values in boldface fonts are those used for the correlation analysis 
              described in Section \ref{sec_correlation}.
              When two spectral lines are detected, the corresponding \NHI 
              are labelled as I (narrower line) and II (broader line).}
\tablenotetext{a}{References for GPS classification: \citet{devries1997}, \citet{tingay1997}, 
             \citet{snellen1998}, \citet{stanghellini1998}, \citet{torniainen2007}, and \citet{vermeulen2003};
              references for CSO classification: 
             \citet{augusto2006}, \citet{bondi2004}, \citet{dallacasa1998},  \citet{nagai2006}, \citet{orienti2006},
             \citet{polatidis2003}, \citet{stanghellini1997}, \citet{stanghellini1999}, and 
              \citet{stanghellini2001}.}
\tablenotetext{b}{3$\sigma$ upper limit.}
\tablenotetext{c}{2$\sigma$ upper limit, derived by V03 from the relation: 
        \NHIns$_{,2\sigma} = 1.82\times10^{18}\times T_{\mathrm{s}} \times \tau_{\rm obs,2\sigma}\times \Delta V$, under the assumption 
        $T_{\mathrm{s}}$=100 K, $\Delta V = 100$ km \persec, and $C_{\mathrm f}=1$.}
\tablenotetext{d}{From results on $\tau_{\rm obs,3\sigma}$, using the relation:  
        \NHIns$_{,3\sigma}= 1.823\times10^{18}\times T_{\mathrm{s}} \times \tau_{\rm obs,3\sigma} \times \Delta V$,
         under the assumptions  $T_{\mathrm{s}}$=100 K, $\Delta V = 100$ km \persec, and $C_{\mathrm f}=1$.}
\tablenotetext{e}{From the 2$\sigma$ estimates by V03, we derived the 3$\sigma$ estimates reported in this row.}
\tablenotetext{f}{\NHI values rescaled to $T_{\mathrm s}=100$ K (A10 and P99 assumed $T_{\mathrm s}=8000$ K).}
\tablenotetext{$\dagger$}{This value of \NHI was chosen for the $T_{\rm s}$ estimate discussed in Section \ref{sec_implications}.}
\tablerefs{($\star$): this work; A10: \citet{araya2010}; C98: \citet{carilli1998}; G06: \citet{gupta2006}; M14: \citet{maccagni2014}; 
           M89: \citet{mirabel1989}; M04: \citet{morganti2004}; M13: \citet{morganti2013}; O06: \citet{orienti2006};
           P03: \citet{pihlstroem2003}; P99: \citet{peck1999}; S07: \citet{saikia2007}; v89: \citet{vangorkom1989}; V00: \citet{veron2000}; 
V03: \citet{vermeulen2003}.}
\end{deluxetable*}

\section{\NH estimates}
\label{subsec_NHapp}

Table \ref{tab_NH} lists the names of the 27 GPS/CSOs of the sample described in Section \ref{sec_sample} (Column 1), their redshift (Column 2), the Galactic column density, \NHns$_{\rm , Gal}$, toward the source (Column 3), the local (i.e., source-intrinsic) absorbing column density, \NH, derived from the X-ray spectrum (Column 4), the X-ray photon spectral index (Column 5), 
and the reference for the X-ray data (Column 6).

In this table, the 22 sources of the  correlation sample  and the corresponding \NH estimates that we 
used for the correlation analysis are highlighted with boldface fonts.  
The criteria that we adopted to select the column density estimates highlighted in Column 4 
among all the available estimates are described below.

We took the \NH estimates from the literature. These estimates were derived from the X-ray spectra with different methods.
When the signal-to-noise ratio was good enough to perform a spectral analysis, \NH was derived by fitting a model spectrum, absorbed by both the Galactic and a local (i.e., at redshift $z_{\rm opt}$) gas column density, to the observed X-ray spectrum. 
In the fitting procedure, the Galactic column density parameter, \NHns$_{\rm,Gal}$, was always fixed, whereas the local column density parameter, \NH, and the photon spectral index parameter, $\Gamma$, were left free to vary. 
On the other hand, when  the signal-to-noise ratio was too low to enable a standard spectral analysis, in 
some cases \citep[e.g.,][]{vink2006,siemiginowska2016}, \NH was constrained by performing the above spectral fitting, but  
with $\Gamma$ frozen to a value (or to a range of values) typical for radio galaxies. In other cases \citep[e.g.,][]{tengstrand2009}, 
\NH constraints were derived with a technique based on a comparison of the observed and simulated hardness ratios of the X-ray 
spectra.

All the available \NH estimates, in the form of either values with corresponding 1$\sigma$
uncertainties, upper limits, or lower limits, are reported in Table \ref{tab_NH}, Column 4. 
The remaining parameters of the X-ray spectral fitting, \NHns$_{\rm,Gal}$ and $\Gamma$, are reported in Columns 3 and 5 of the same Table, respectively.

For some of the sources, different estimates of \NH were derived from the analysis of X-ray observations carried out at different epochs and/or with different detectors, as well as from 
the analysis of the same X-ray data set either by different authors or with different techniques. 
Furthermore, the X-ray spectra of some sources could be satisfactorily fit with different models, implying different absorption estimates.

In all the cases in which we had multiple choices for the \NH estimate, we adopted the following criteria for the purpose of our correlation analysis, unless otherwise stated. 
We discarded the values derived from soft X-ray data only (e.g., ROSAT data).
When more recent X-ray data of a given source proved that the older results were affected by poor angular resolution,  
we chose the most accurate (also the most recent) value when the two values were consistent with each other at the 1$\sigma$ level,
whereas we took both values into account when they were not consistent with each other, 
in that we could not rule out long-term column-density variations.
When two different data sets yielded different \NH upper limits, we chose 
the less stringent one.\footnote{We note that this choice differs from the choice taken for \NHI upper limits at the end of Section \ref{subsec_NHI}: when we have upper limits, we choose the most and the least stringent limit for \NHI and \NHns, respectively. This choice is motivated by the fact that \NHI upper limits are derived from the continuum flux density and the rms noise level of the radio spectra, whereas \NH upper limits are derived from the likelihood function of the parametric fit to the X-ray spectra. Therefore, in the \NHI case, the most stringent upper limit corresponds to data affected by the lowest noise, whereas in the \NH case, the quality of the data of different observations is comparable and the least stringent upper limit represents the most conservative choice.}
When the same data set was analyzed by different authors with the same model, we chose the result of the most recent analysis. 
When the same data set was analyzed by the same authors with the same model, but with different techniques, we chose the result of the most robust analysis (e.g., we discarded an \NH value derived by fixing the spectral index $\Gamma$ in favor of an upper or lower limit to \NH derived without any constraints on $\Gamma$). 
When more than one model spectrum could be satisfactorily fit to a given observed spectrum, we selected the \NH estimate derived from the simplest model among those that include a Compton-thin absorber. When a Compton-thick absorber scenario could also apply to the source,
we always took into account also the \NH lower limit (\NH $\gtsim 10^{24}$ \percmq)
derived in the framework of this scenario.
The latter case applied to a subsample of three sources.
Finally, when a {\it lower} limit to \NH was derived in a Compton-thin scenario,\footnote{This case applies to one source only, i.e. 0108$+$388.}, we could not state whether the source 
was Compton-thin or Compton-thick; we thus considered both scenarios to be plausible, as in the previous case.
In the Compton-thin scenario, we associated with the source 
an \NH value equal to the mean of the 3$\sigma$ lower limit and the physical upper bound of 
the Compton-thin \NH range (i.e., \NH $\simeq 10^{24}$ \percmq).
We considered the \NH range as a $\pm 3\sigma$ interval, and thus associated with the mean \NH the corresponding 1$\sigma$ uncertainty.
Overall, the ambiguity between a Compton-thin and a Compton-thick absorber (and between the corresponding values of \NHns) 
affected a subsample of four sources of the correlation sample.

\LongTables 
\begin{deluxetable*}{lccccc}
\tabletypesize{\scriptsize}
\tablecaption{X-ray Column Densities and Spectral Parameters. \label{tab_NH}}
\tablewidth{0pt}
\tablehead{
\colhead{Source Name} &
\colhead{$z_{\mathrm {opt}}$} &
\colhead{\NHns$_{\rm,Gal}$} &
\colhead{\NH} &     
\colhead{$\Gamma$} & 
\colhead{References} \\
\colhead{B1950} &
\colhead{~~} &       
\colhead{($10^{20}$ \percmq)} & 
\colhead{($10^{22}$ \percmq)} & 
\colhead{~~}  &
\colhead{~~} \\ 
\colhead{(1)} &
\colhead{(2)} &
\colhead{(3)} &
\colhead{(4)} &
\colhead{(5)} &
\colhead{(6)} \\
}
\startdata
{\bf 0019$-$000} 	& 
0.305   		& 
2.7                 	& 
$\mathbf{<100}$        	& 
\nodata               	 & 
T09  \\ 
0026$+$346 & 
0.517     & 
5.6               &  
$1.0^{+0.5}_{-0.4}$            & 
$1.43^{+0.20}_{-0.19}$   & 
G06 \\     
{\bf 0035$+$227}   & 
$0.096\pm0.002$   & 
3.37    & 
$\mathbf{1.4^{+0.8}_{-0.6}}$    & 
1.7\tnm{a,b}        & 
S16 \\
{\bf 0108$+$388} & 
0.66847   & 
5.8                 & 
$57\pm20$                     & 
1.75\tnm{a}       & 
V06 \\	
~~~        & 
~~~       & 
5.8                 & 
$>18$\tnm{c}       & 
\nodata                  & 
V06 \\	
~~~        & 
~~~       & 
~~~                 & 
$\mathbf{>5}$        & 
\nodata                  & 
T09 \\	
0116$+$319 & 
0.06      &  
5.67               &  
\nodata  	             & 
\nodata                  & 
S16 \\             	
{\bf 0428$+$205} & 
0.219   & 
19.6            & 
$\mathbf{<0.69}$\tnm{d}     & 
$[0.63-2.62]$\tnm{a}    & 
T09  \\ 
{\bf 0500$+$019} & 
0.58457 & 
8.3             & 
$0.5^{+0.3}_{-0.2}$           & 
$1.62^{+0.21}_{-0.19}$   & 
G06 \\ 
~~~        & 
~~~       & 
\nodata             & 
$\mathbf{0.5^{+0.18}_{-0.16}}$ & 
$1.61^{+0.16}_{-0.15}$  & 
T09 \\
0710$+$439 & 
0.518     & 
8.11                & 
$0.44\pm0.08$                 & 
$1.59\pm0.06$            & 
V06 \\
~~~        & 
~~~       & 
~~~                 & 
$0.51^{+0.17}_{-0.14}$        & 
$1.59\pm0.06$            & 
T09 \\	
~~~        & 
~~~       & 
8.0                 & 
$0.58\pm0.08$	             & 
$1.59\pm0.07$            & 
S16 \\
~~~        & 
~~~       & 
8.0                 & 
$0.61^{+0.04}_{-0.08}$	     & 
$1.64^{+0.09}_{-0.07}$   & 
S16 \\	
~~~        & 
~~~       & 
8.0                 & 
$0.49^{+0.07}_{-0.06}$	     & 
\nodata                  & 
S16 \\	
~~~        & 
~~~       & 
8.0                 & 
$0.58\pm0.08$   		     & 
$1.42^{+0.14}_{-0.20}$   & 
S16 \\	
~~~        & 
~~~       & 
8.0                 & 
$0.56\pm0.08$   		     & 
$1.39^{+0.17}_{-0.23}$   & 
S16 \\	
~~~        & 
~~~       & 
8.0                 & 
$1.02^{+0.29}_{-0.22}$	     & 
$1.75^{+0.11}_{-0.10}$   & 
S16 \\	
{\bf 0941$-$080} & 
0.2281$\pm$0.0013  & 
3.7  & \nodata                       & 
2\tnm{a}                 & 
G06 \\ 
~~~        & 
~~~       & 
3.67                & 
\nodata                       & 
$2.62^{+1.29}_{-1.03}$   & 
S08 \\
~~~        & 
~~~       & 
3.67                & 
$\mathbf{<1.26}$\tnm{c}       &
 $2.28^{+0.67}_{-0.61}$   & 
 S08, O10 \\ 
~~~        & 
~~~       & 
3.67                & 
$<0.53$\tnm{c}                & 
$[1.7-1.9]$\tnm{a}       & 
S08, O10\\
~~~        & 
~~~       & 
3.67                & 
$<100$                        & 
\nodata                  &
T09 \\
{\bf 1031$+$567} & 
0.45   & 
0.56             & 
$\mathbf{0.50\pm0.18}$        & 
1.75\tnm{a}              & 
V06 \\  
~~~              & 
~~~    & 
\nodata          & 
$0.50\pm0.18$                 & 
\nodata                  & 
T09\\			    
{\bf 1117$+$146} & 
0.362   & 
2.0             & 
$\mathbf{<0.16}$\tnm{d}     &
$[0.63-2.62]$\tnm{a}     & 
T09  \\ 
1245$+$676 & 
0.107   & 
\nodata             & 
\nodata     &
\nodata     & 
S16, W09  \\ 
{\bf 1323$+$321} & 
0.370 & 
1.2             & 
$\mathbf{0.12^{+0.06}_{-0.05}}$ & 
$1.7\pm0.2$            & 
T09 \\                                                  
{\bf 1345$+$125} & 
0.12174 & 
1.1             & 
$4.2^{+4.0}_{-2.4}$            & 
$1.6^{+1.2}_{-0.8}$      & 
O00 \\ 
~~~        & 
~~~     & 
\nodata               & 
$\mathbf{4.8\pm0.4}$           & 
$1.1^{+0.7}_{-0.8}$      & 
T09\\
~~~        & 
~~~     & 
1.9                   & 
$\mathbf{2.543^{+0.636}_{-0.580}}$ & 
$1.10^{+0.29}_{-0.28}$ & 
S08 \\ 
~~~        & 
~~~     & 
1.9                   & 
$2.28^{+0.36}_{-0.35}$\tnm{e}  & 
$1.27^{+0.21}_{-0.19}$   & 
S08 \\ 
{\bf 1358$+$624} & 
0.431   & 
1.96            & 
$3.0\pm0.7$                    &
$1.24\pm0.17$            & 
V06 \\
~~~	   & 
~~~     & 
\nodata               & 
$\mathbf{2.9^{+2}_{-1}}$       & 
$1.24\pm0.17$            & 
T09 \\ 
{\bf 1404$+$286} & 
0.07658 & 
1.4             & 
$\mathbf{0.13^{+0.12}_{-0.10}}$\tnm{e} & 
$2.1^{+0.6}_{-0.3}$\tnm{e}, $0.7^{+0.3}_{-0.4}$\tnm{f} & 
G04 \\        
~~~	   & 
~~~     & 
1.4                   & 
$0.12^{+0.09}_{-0.08}$\tnm{e} , $24.0^{+10.0}_{-8.0}$\tnm{f,g} & 
$2.0^{+0.4}_{-0.3}$      & 
G04 \\ 
~~~        &
 ~~~     & 
1.4                   & 
$0.11\pm0.05$\tnm{e}, $\mathbf{>90}$\tnm{f,g} & 
$2.21^{+0.19}_{-0.14}$\tnm{h}              & 
G04 \\
~~~        & 
~~~     & 
1.4                   & 
$0.19^{+0.13}_{-0.10}$\tnm{e}, $>90$\tnm{f,g} & 
$2.6\pm 0.5$\tnm{e}, 2\tnm{a,f}   & 
G04 \\
~~~        & 
~~~     & 
1.4                   & 
$0.09^{+0.08}_{-0.06}$\tnm{e}, $>90$\tnm{f,g} & 
$2.2\pm 0.4$\tnm{h}      	 & 
G04 \\
~~~        & 
~~~     & 
1.4                   & 
$<0.08$\tnm{e}, $>90$\tnm{f,g} & 
$1.2^{+2.0}_{-0.3}$                              & 
G04 \\ 
~~~        & 
~~~     & 
\nodata               & 
$0.08^{+0.31}_{-0.05}$\tnm{e} & 
$2.4^{+2.1}_{-1.1}$                               & 
G04 \\ 
~~~        & 
~~~     & 
\nodata               & 
$>90$                         & 
$2.21^{+0.19}_{0.14}$                             & 
T09 \\ 
{\bf 1509$+$054} & 
0.084 & 
3.29               & 
$\mathbf{<0.23}$ & 
$1.0\pm0.2$                                                    & 
S16, K09 \\
{\bf 1607$+$268} & 
0.473   & 
3.8             & 
$<0.2$                         & 
$0.4\pm 0.3$             & 
T09   \\ 
~~~        & 
~~~     & 
3.8                   & 
$>60$                          & 
\nodata                  & 
T09   \\
~~~        & 
~~~     & 
4.1                   & 
$\mathbf{<0.18}$               & 
$1.4\pm 0.1$             & 
S16 \\
{\bf 1718$-$649} &  
0.0142  &  
7.15                & 
$\mathbf{0.08\pm 0.07}$        &  
$1.6\pm0.2$       & 
S16   \\                   
{\bf 1843$+$356}   & 
0.764 & 
6.75                  & 
$\mathbf{0.8^{+0.9}_{-0.7}}$   & 
1.7\tnm{a,b}         & 
S16 \\
{\bf 1934$-$638}  & 
0.18129 & 
\nodata              & 
$\mathbf{>250}$   	      & 
$1.9^{+0.5}_{-0.6}$      & 
R03  \\ 
~~~         & 
~~~     & 
\nodata              & 
$<2.0$                         & 
\nodata                 & 
R03  \\ 
~~~         & 
~~~     & 
6.16                 & 
$\mathbf{0.08^{+0.07}_{-0.06}}$ & $1.67^{+0.15}_{-0.16}$ & 
S16  \\ 
{\bf 1943$+$546} & 
0.263   & 
13.15                 & 
$\mathbf{1.1\pm0.7}$           & 
1.7\tnm{a,b}        & 
S16  \\
{\bf 1946$+$708} & 
0.101   & 
\nodata       & 
$2.60^{+2.5}_{-1.90}$    & 
$2.6^{+0.6}_{-0.7}$      & 
R03  \\  
~~~        & 
~~~     & 
\nodata             & 
$\mathbf{>280}$                  & 
\nodata                  & 
R03  \\  
~~~        & 
~~~     & 
8.57                & 
$\mathbf{1.7^{+0.5}_{-0.4}}$     & 
$1.7\pm0.4$              & 
S16 \\   
2008$-$068 & 
0.547   & 
5.0                 & 
$<0.48$\tnm{d}                 & 
$[0.63-2.62]$\tnm{a}     & 
T09   \\ 
{\bf 2021$+$614} & 
0.227   & 
14.01         & 
$\mathbf{<1.02}$                 & 
$0.8^{+0.3}_{-0.2}$        & 
S16 \\
{\bf 2128$+$048} & 
0.99   & 
5.2             & 
$0.3^{+0.81}_{-0.3}$            & 
$1.5^{+0.6}_{-0.7}$        & 
G06 \\       
~~~        & 
~~~    & 
5.2                   & 
\nodata                         & 
$1.28^{+0.42}_{-0.41}$     & 
S08 \\ 
~~~        & 
~~~    & 
5.2                   & 
$<1.56$                         & 
$1.44^{+0.89}_{-0.70}$     & 
S08, O10 \\ 
~~~        & 
~~~    & 
5.0                   & 
$\mathbf{<1.9}$                 & 
$1.98^{+0.5}_{-0.4}$       & 
T09 \\
{\bf 2352$+$495} & 
0.2379 & 
12.4            & 
$0.66\pm0.27$                 & 
1.75\tnm{a}          & 
V06  \\      
~~~        & 
~~~    & 
\nodata               & 
$\mathbf{4^{+7}_{-3}}$        & 
$1.8^{+1.6}_{-0.9}$        & 
T09 \\  
\enddata
\tablenotetext{}{{\bf Note. }The values of \NH in boldface fonts are those used for the correlation analysis in Section \ref{sec_correlation}.}
\tablenotetext{a}{Fixed.}
\tablenotetext{b}{By varying $\Gamma$ in the range [1.4-2.0], the authors get \NH values consistent, to within 1$\sigma$, with the \NH value given in this Table.}
\tablenotetext{c}{$3\sigma$ (upper or lower) limit.} 
\tablenotetext{d}{Upper bound of the 1.6$\sigma$ interval [\NHns$_{min}$, \NHns$_{max}$] 
     about the curve corresponding to the nominal hardness ratio, when $\Gamma$ is constrained to the given range.} 
\tablenotetext{e}{Soft X-rays.}
\tablenotetext{f}{Hard X-rays.} 
\tablenotetext{g}{\NHns$^{\mathrm{hard}}>9\times10^{23}$ \percmq is associated with a Compton-reflection model for the hard X-rays.}
\tablenotetext{h}{Intrinsic, for the Compton-reflection model.}
\tablerefs{G04: \citet{guainazzi2004}; G06: \citet{guainazzi2006};  K09: \citet{kuraszkiewicz2009}; O00: \citet{odea2000}; O10: \citet{ostorero2010}; 
           R03: \citet{risaliti2003}; S08: \citet{siemiginowska2008}; S16: \citet{siemiginowska2016}; T09: \citet{tengstrand2009}; V06: \citet{vink2006}; 
           W09: \citet{watson2009}.}
\end{deluxetable*}

~~\\

\bibliographystyle{apj}
\bibliography{msbib}

\end{document}